\documentclass [manuscript]{aastex}
\usepackage{amsmath}

\shorttitle{REDUCED MASS LOSS RATES AND ANISOTROPIC WINDS} \shortauthors{Guo}

\begin{document}


\title{ESCAPING PARTICLE FLUXES IN THE
ATMOSPHERES OF CLOSE-IN EXOPLANETS. II. REDUCED MASS LOSS RATES AND ANISOTROPIC WINDS }


\author{J. H. Guo\altaffilmark{1}}
\affil{National Astronomical Observatories/Yunnan Observatory,
Chinese Academy of Sciences, P.O. Box 110, Kunming 650011, China; guojh@ynao.ac.cn}



\altaffiltext{1}{Key Laboratory for the Structure and Evolution of
Celestial Objects, CAS, Kunming 650011, China}


\begin{abstract}
In Paper I, we presented a one-dimensional hydrodynamic model for the
winds of close-in exoplanets. However, close-in exoplanets are
tidally locked and
irradiated only on the day sides by their host stars. This requires
two-dimensional hydrodynamic models with self-consistent radiative
transfer calculations. In this paper, for the tidal locking
(two-dimensional radiative transfer) and non-tidal locking cases
(one-dimensional radiative transfer), we constructed a multi-fluid two-dimensional hydrodynamic
model with detailed radiative transfer to depict the escape of particles. We found that the
tidal forces (the sum of tidal gravity of the star and centrifugal force
due to the planetary rotation) supply significant accelerations and result in anisotropic winds.
An important effect of the tidal
forces is that it severely depresses the outflow of particles near the polar
regions where the density and the radial velocity are a factor of few-ten smaller
than those of the low-latitude regions. As a consequence, most particles escape the surface of the
planet from the regions of low-latitude. Comparing the tidal and
non-tidal locking cases, we found that their
optical depths are very different so that the flows also emerge with
a different pattern. In the case of
the non-tidal locking, the radial velocities at the base of the wind are  higher than the meridional velocities.
However, in the case of tidal locking, the meridional velocities dominate the flow at the base of the wind, and they can
transfer effectively mass and energy from the day sides to the night
sides. Further, we also found that the differences of the winds show middle extent at large radii.
It expresses that the structure of the wind at the base can be changed by the two-dimensional radiative transfer
due to large optical depths, but the
extent is reduced with an increase in radius. Because the escape is depressed in the polar regions, the mass loss rate
predicted by the non-tidal locking model,
in the order of magnitude of $10^{10}$ g s$^{-1}$, is a factor of 2 lower than that
predicted by one-dimensional hydrodynamic model. The results of
tidal locking show that the mass loss rate is
decreased a order of magnitude, only 4.3$\times$ $10^{9}$ g s$^{-1}$,
due to large optical depths on the night side. 
We also found that the distributions of hydrogen atoms show clear variations from the day side to night side,
thus the origin of the excess absorption in Ly $\alpha$
should be reexamined by using multi-dimensional hydrodynamic models.

\end{abstract}





\section{INTRODUCTION}

The excess absorption in Lyman $\alpha$ of HD 209458b and HD 189733b
have been found by Vidal-Madjar et al. (2003) and Lecavelier des
Etangs et al. (2010). From the excess absorption those authors
suggested that the hydrogen clouds of HD 209458b and HD 189733b are
above their Roche lobe. Most of the explanations attribute the
excess absorption to mass loss from the surfaces of exoplanets
(Lammer et al. 2003; Lecavelier des Etangs et al. 2004; Yelle 2004,
2006; Tian et al. 2005; Garc\'{\i}a Mu\~{n}oz 2007; Penz et al.
2008; Murray-clay et al. 2009; Lammer et al. 2009; Guo 2011 (Paper I)).
To simplify the calculations, all the hydrodynamic models above are one dimensional.
Recently, two- and three-dimensional
hydrodynamic models have been presented by Stone \& Proga
(2009) and Schneiter et al. (2007), however, some challenging physical processes have
been neglected by them.
All the above one-dimensional models described thermal particle escape, namely, the particles of the planets
escape from their gravity potential wells as a result of
the irradiation from the host star. However, the charge
exchange between the stellar wind and the planetary escaping
exosphere can result in the same observation phenomenons (Holmstr\"{o}m
et al. 2008). The loss of nonthermal neutral atoms via interactions
between the stellar wind and the exosphere have been discussed by
many researchers (Erkaev et al. 2005; Holmstr\"{o}m et al. 2008; Ekenb\"{a}ck et al.
2010). Their results showed that the fresh neutral atoms produced by the
charge exchange can match the Ly $\alpha$ excess absorption.

Although current theoretic models predicted the mass loss at the upper
atmosphere of Hot Jupiters, no clear observational evidences
support the thermal or non-thermal explanations.
Either the energetic HI of stellar origin or thermal HI populations
in the planetary atmosphere can fit the observations of Ly $\alpha$
(Ben-Jaffel \& Hosseini 2010). Moreover, Koskinen et al. (2010)
found that the excess absorption can be explained solely by
absorption in the upper atmosphere and the process of charge
exchange might not be necessary. According to the examples from
our solar system, however, the charge exchange can play an important role in
explaining the excess Ly $\alpha$ absorption (see Lammer
2011 for more details).

The extent of understanding the excess absorption depends strongly
on the physics described in theoretical models. To fully distinguish the thermal
or non-thermal processes, the single-fluid and one-dimensional
models are not enough. Yelle (2004) and Garc\'{\i}a Mu\~{n}oz (2007) have
used the diffusion approximation to depict the velocity differences
between different species, and photochemistry, ionization and recombination
have also been included. Moreover, Guo (2011) presented a
multi-fluid one-dimensional model to depict the atmospheric escape
and predicted reasonable mass loss rates for HD 209458b and HD
189733b. The multi-fluid model is more advantageous than
single-fluid one because the interactions among different species
between the stellar and planetary winds are depicted by its own
continuity, momentum and energy equations. 
However, the model of Paper I can be improved in two points at least.
First, the one-dimensional hydrodynamic model only simulates the case
along the line connecting the center of planet and the center of the star. 
For the close-in expolanets, the tidal forces between the
star and planet tend to circularize the orbit and synchronize
the rotation of the planet with the orbital period (Guillot et al.
1996; Trilling 2000). Thus, for the close-in exoplanets it is reasonable
to assume that their same sides are always facing the star. If the
irradiation and tidal effects of the host star are included,
multi-dimensional models are more suitable. Second, the constraint
of stellar wind has been omitted. Stone \& Proga (2009) discussed
the interaction between the stellar and planetary winds and found
that the planetary wind is confined to a small volume by the stellar
wind. However, their model used the assumption of a single-fluid so
the processes of charge exchange could not be simulated.

These issues as above motivated us to construct a two-dimensional
hydrodynamic model with realistic treatment of radiative transfer and tidal forces.
We have developed a time-independent model with the
one-dimensional assumption, however, it is impractical to extend the
method to two-dimension because of the existence of critical points.
In this paper we constructed a time-dependent two-dimensional model
including the micro-physics processes, the influence of
tide and radiative transfer. The interactions between the stellar and
planetary winds will be discussed in the further work.

This paper aims to calculate the escape of atomic hydrogen
and protons through the solution to their mass, momentum and energy
equations. The microscopic physics processes are covered
in the mass, but not in the momentum and energy equations (Section 2.1). The processes
of radiative transfer are discussed in Section 2.2. We used
semi-implicit scheme to calculate these equations (Section 2.3) and
tested the model in Section 3. We incorporated one- and two-dimensional
radiative transfer into the
hydrodynamic models (Section 4), and focus on the differences
between them. In Section 5 we discussed the influence of the boundary conditions
and stellar activity. Our
results are summarized in Section.6

\section{THE MODEL}

\subsection{Equations and Assumptions}
In Paper I (Guo 2011) we described a steady-state, radial
expansion of plasma containing three species: atomic hydrogen (h),
protons (p) and electrons (e). In this paper, we extend the previous
work to two-dimensions. But thermal electrons are not included.
Atomic hydrogen (h) and
protons (p) have their own continuity equations
and are described by a particle density $n_{h}$ and $n_{p}$.
Since this model deals with a mixture of atomic hydrogen and
protons, the following processes are considered: photoionization and
recombination. Further, we assumed that the velocities and temperatures of the particles are the same,
namely, $u_{h}=u_{p}=u$ and $T_{h}=T_{p}=T$.
The assumption is acceptable because the mass loss rates predicted by this model
are higher than 10$^{9}$ g s$^{-1}$ (see Section 3 and 4), if the mass loss rates are below
the value the velocities and temperature should be calculated separately because of the decoupling of particles (Guo 2011).
As in Paper I, we did not
include H$_{2}$ in this model because the thermosphere of close-in
planets should be composed primarily of H and H$^{+}$. The location
of transition from H$_{2}$ to H is about 1.1$R_{P}$ (Yelle 2004).

We described the planetary wind by using multi-particle two-dimensional
hydrodynamic equations that can be written as
\begin{equation}
\frac{\partial n_{j}}{\partial t}+\triangledown \cdot (n_{j}\mathbf{u})=S_{j}
\end{equation}

\begin{equation}
\frac{\partial (n\mathbf{u})}{\partial t}+\triangledown \cdot (n\mathbf{u}\mathbf{u})+\triangledown p=n \mathbf{a}_{ext}
\end{equation}

\begin{equation}
\frac{\partial [n(e+\frac{{u}^{2}}{2})]}{\partial t}+\triangledown \cdot [n\mathbf{u}(e+\frac{{u}^{2}}{2})]+\triangledown (p\mathbf{u})=n \mathbf{a}_{ext}\cdot\mathbf{u}+Q,
\end{equation}
where $n_{j}$(\emph{j=h,p}) is particle number density of hydrogen
atoms and ions. n=n$_{h}$+n$_{p}$ is the number density of the gas as
a whole. $S_{j}$ denotes the production/destruction of particles.
The specific internal energy is
$e=\frac{\mathcal{R}T}{\gamma-1}$, and
$u^{2}=u_{r}^{2}+u_{\theta}^{2}$ is the sum of squares of
velocities, where $\hat{r}$ and $\hat{\theta}$ velocities are
$u_{r}$ and $u_{\theta}$. $\gamma$ is set to $5/3$. The accelerations produced by the gravities
of planet and star as well as the centrifugal forces due to the rotation of the planet
around the star are included in the term of $\mathbf{a}_{ext}$. The net heating rate is expressed as
$Q=H-L$, where H is the heating from irradiation of the star and L is cooling due to the atmospheric radiation.

The hydrodynamic equations are solved in spherical polar (r,$\theta$)
coordinates. Here $\theta$=0 is the direction towards the star, and $\theta$=$\pi$ is the direction away from the star. 
In conservation form, these equations can be expressed as
\begin{equation}
\frac{\partial Q}{\partial t}+\frac{\partial F}{\partial r}+\frac{\partial G}{\partial \theta}=S_{1}+S_{2},
\end{equation}

where \emph{Q} is the state vector, \emph{F} and \emph{G} are the flux
vectors, $S_{1}$ denotes the contributions from the planetary gravity
and the tidal forces of stars, and $S_{2}$ includes the source/sink terms
that are relative with photoionization/recombination and heating/cooling.

The state vector Q are defined as
\begin{align}
    Q=\left[
           \begin{array}{ccc}
           Q_{1}\\
           Q_{2}\\
           Q_{3}\\
           Q_{4}\\
           Q_{5}\\
           \end{array}
           \right]
      =\left[
           \begin{array}{ccc}
             n_{h}r^{2} \sin \theta \\
             n_{p}r^{2} \sin \theta \\
             (n_{h}+n_{p})r^{2} \sin \theta u_{r}\\
             (n_{h}+n_{p})r^{2} \sin \theta u_{\theta}\\
             (n_{h}+n_{p})r^{2} \sin \theta(e+\frac{u^{2}}{2}).\\
           \end{array}
         \right]
 \end{align}

The flux vector [\emph{F, G}] and the
source terms $S_{1}$ are defined as
\begin{align}
    F=\left[
           \begin{array}{ccc}
             \frac{Q_{1}Q_{3}}{Q_{1}+Q_{2}} \\
             \frac{Q_{2}Q_{3}}{Q_{1}+Q_{2}} \\
             \frac{Q_{3}^{2}}{Q_{1}+Q_{2}}+a^{2}(Q_{1}+Q_{2})\\
             \frac{Q_{3}Q_{4}}{Q_{1}+Q_{2}}\\
             \frac{Q_{3}Q_{5}}{Q_{1}+Q_{2}}+a^{2}Q_{3}\\
           \end{array}
         \right]
 \end{align}

\begin{align}
    G=\left[
           \begin{array}{ccc}
             \frac{Q_{1}Q_{4}}{r(Q_{1}+Q_{2})} \\
             \frac{Q_{2}Q_{4}}{r(Q_{1}+Q_{2})} \\
             \frac{Q_{3}Q_{4}}{r(Q_{1}+Q_{2})}\\
             \frac{Q_{4}^{2}}{r(Q_{1}+Q_{2})}+\frac{a^{2}(Q_{1}+Q_{2})}{r}\\
             \frac{Q_{4}Q_{5}}{r(Q_{1}+Q_{2})}+\frac{a^{2}Q_{4}}{r}\\
           \end{array}
         \right]
 \end{align}

\begin{align}
    S_{1}=\left[
           \begin{array}{ccc}
             0 \\
             0 \\
             2a^{2}\frac{(Q_{1}+Q_{2})}{r}+\frac{Q_{4}^{2}}{r(Q_{1}+Q_{2})}+(Q_{1}+Q_{2})a_{r} \\
             a^{2}\frac{(Q_{1}+Q_{2})\cos \theta}{r\sin \theta}-\frac{Q_{3}Q_{4}}{r(Q_{1}+Q_{2})}+(Q_{1}+Q_{2})a_{\theta}\\
             Q_{3}a_{r}+Q_{4}a_{\theta}\\
           \end{array}
         \right],
 \end{align}
where $a=\sqrt{\mathcal{R} T}$ is the isothermal sound velocity, and
$a_{r}$ and $a_{\theta}$ are the accelerations in $\hat{r}$ and
$\hat{\theta}$ directions, respectively. The accelerations are given
by
\begin{equation}
a_{r}=-\frac{\partial U}{\partial r},
\end{equation}

\begin{equation}
a_{\theta}=-\frac{1}{r}\frac{\partial U}{\partial \theta}.
\end{equation}

For the position vector $\mathbf{x}=(r,\theta)$, the effective
potential $U$ is given by

\begin{equation}
U({\mathbf{x}})=-\frac{GM_{p}}{|\mathbf{x}|}-\frac{GM_{p}}{|\mathbf{x}-D|}-\frac{1}{2}|\Omega\times \mathbf{x}|^{2},
\end{equation}
where D is the distance of star,
$\Omega=[G(M_{\ast}+M_{p})/D^{3}]^{1/2}$ is the angular velocity of
orbital plane, and $M_{*}$ and $M_{p}$ are the stellar and planetary
mass, respectively. Finally, we can obtain the accelerations as
\begin{equation}
a_{r}=-\frac{GM_{p}}{r^{2}}+\frac{GM_{\ast}(D\cos \theta-r)}{(D^{2}+r^{2}-2Dr\cos \theta)^{3/2}}-G\frac{M_{\ast}}{D^{3}}\cos \theta(D-r\cos \theta)
\end{equation}

\begin{equation}
a_{\theta}=-\frac{GM_{\ast}D\sin \theta}{(D^{2}+r^{2}-2Dr\cos \theta)^{3/2}}+G\frac{M_{\ast}}{D^{3}}\sin \theta(D-r\cos \theta).
\end{equation}

In Equation (12), the first term represents the gravity of planet. The second
and third terms as well as the two terms of Equation (13) are the accelerations produced by the gravity of host
stars and orbital motion, respectively. Except the first
term of Equation (12), we refer to the other terms as "tidal accelerations". How do the tidal accelerations alter the behaviors of flows?
We can estimate this by considering some special cases. In the case of $\theta$ =0$^{\circ}$, Equation (12) can be simplified
to $a_{r}=-\frac{GM_{p}}{r^{2}}+\frac{GM_{\ast}}{(D-r)^{2}}-G\frac{M_{\ast}}{D^{3}}(D-r)$. This form
is consisitent with the Equation (15) of Garc\'{\i}a Mu\~{n}oz (2007) and hints that the tidal forces supply positive acceleration along
the line connecting the planet to the star. However, the radial tidal acceleration gradually decreases with the increase of $\theta$ and become negative
at 60$^{\circ}$ $\lesssim \theta  \lesssim$ 120$^{\circ}$. This means that the escape of particles in the polar regions can be depressed
by the tidal forces. In addition, we also note that $\hat{a_{\theta}}$ is always negative (positive) if the $\theta$ is smaller (greater) than $\sim$ 90$^{\circ}$.
These characteristics will also be discussed further in Section 4.

The Coriolis forces have been neglected in the calculations.
In the regions close to host stars, the escape velocities of particles which are in the order of magnitude
of 20-40 km s$^{-1}$ are smaller than the orbital velocities (in the order
of magnitude of 100 km s$^{-1}$). This means that the planetary wind can be deflected by
Coriolis forces. However,
our calculations are limited to the regions close to exoplanets
where the velocity of the wind is much less than the orbital velocity.
Thus, Coriolis forces can be neglected safely near the planet.

The source terms $S_{2}$ can be expressed as
\begin{align}
    S_{2}=\left[
           \begin{array}{ccc}
             \frac{Q_{2}^{2}}{r^{2}\sin \theta}\gamma _{rec}-Q_{1} \gamma _{pho} \\
             Q_{1}\gamma _{pho}-\frac{Q_{2}^{2}}{r^{2}\sin \theta} \gamma _{rec}  \\
             0 \\
             0 \\
             r^{2}\sin \theta(H-L)/m_{H}\\
           \end{array}
         \right],
 \end{align}
where m$_{H}$ is the mass of hydrogen atom, $\gamma_{\mathrm{pho}}=\sum_{\nu}\frac{F_{\nu}e^{-\tau_{\nu}}}{h\nu}\sigma_{\nu}$
s$^{-1}$ (where $F_{\nu}$ is photon energy at frequency $\nu$,
$\tau_{\nu}$ is the optical depth at frequency $\nu$, the cross section of photoionization of
hydrogen is $\sigma_{\nu}=6\times10^{-18}(h\nu/13.6eV)^{-3}$ cm$^{2}$ (Murray-Clay et al. 2009)) and
$\gamma_{\mathrm{rec}}=2.7\times10^{-13}(\frac{T}{10^{4}})^{-0.9}$
cm$^{-3}$s$^{-1}$ (Murray-Clay et al. 2009) represent the photoionization and
recombination rates, respectively. In the continuity equation, the sources and sinks to the particle flux density are due to
photoionization and recombination. In energy equation, the sources and sinks are heating and cooling, respectively.

The heating processes in the mixed flows are complex.
The models of Murray-Clay et al. (2009) and Paper I in which
the heating from the host stars is simplified to a single photon
of 20 $\textmd{ev}$ predicted reasonable mass loss rates.
However, the assumption of single photon energy could miss some important physical
processes, such as the penetration of higher energy photons deeper
in the atmosphere. In fact, the stellar radiation ionizes the species to produce high-energy
photoelectrons that share their energy with other species via
collisions. The heating of exoplanet atmospheres mainly comes from
photoelectrons produced by photoionization, therefore,
here we included EUV
irradiance shortward of 912 ${\AA}$, namely, the threshold value of photoionization
in hydrogen atoms.

In an ionized atmosphere composed of H and H$^{+}$, photoelectrons distribute
their energy via the collision with thermal electrons. H$^{+}$ obtain energy from
thermal electrons via Coulomb collisions while hydrogen atoms share the energy of
thermal electrons by elastic and inelastic collisions. Here we assumed that H and H$^{+}$ have the same temperatures,
therefore, the total heating of the atmosphere
can be written as
\begin{equation}
H=n_{h}\sum_{\nu}\eta_{\nu} F_{\nu} e^{-\tau_{\nu}}\sigma_{\nu}.
\end{equation}

For different photon energies,
we defined the net heating efficiency as $\eta_{\nu}=(h\nu-13.6)/h\nu$ that is the maximum heating
efficiency. In fact, the fraction of photon energy deposited as heat in the atmosphere
of exoplanet should be smaller than the definition above. It also shows that the heating
efficiency is higher for the photons with higher energy.
Equation (15) hints
that the contributions of heating by lower energy photons are comparable
with those of higher energy photons because the heating of atmosphere is proportional to
heating efficiency, but is inversely with cube of cross section of photoionization.

The cooling rate adopted in the models of Murray-Clay et al. (2009) and Paper I assumed that the atmosphere
is optically thin in the Ly $\alpha$ line. Murray-Clay et al. (2009) have also
validated that a photon can diffuse out of the wind in a typical mass loss
rate of 10$^{10}$ g s$^{-1}$ (see Appendix C of Murray-Clay et al. 2009).
However, the observations indicate that the atmosphere is optically
thick even in the wings of of Ly $\alpha$. Therefore, it is uncertain if the Ly $\alpha$
cooling should be included. Here we included the radiative cooling by recombination and free-free radiation, namely,

\begin{equation}
L=L_{rec}+L_{ff}  \mathrm{erg cm^{-3}s^{-1}}.
\end{equation}

In addition, we can not calculate the infrared emission of H$^{+}_{3}$ as done by
Yelle (2004) and Garc\'{\i}a Mu\~{n}oz (2007) due to the lack of photochemistry
in our model. The cooling of H$^{+}_{3}$
could be important because it can modify the location of maximum temperature, but the mass loss
rates are not affected by the cooling of H$^{+}_{3}$ (Penz et al. 2008).

There are not resolved emission spectra from the host stars of exoplanets.
For HD 209458, we used the EUV spectra of the sun as its emission spectra.
Those data are taken from Richards et al. (1994). Our calculations are completed
with a solar proxy $P_{10.7}=80$ that corresponds to a middle-low level
of stellar activity.

\subsection{Radiative transfer}
The radiative transfers are the most important processes in determining
the atmospheric structure of planet. In the case of non-tidal locking, both hemispheres 
of the planet can receive the irradiation from its host star. Thus, the one-dimensional radiative transfer
can approximately depict the process. In the assumption of 1D radiative transfer, the optical
depth at frequency $\nu$ can be written as
\begin{equation}
\tau_{\nu}=\int_{R_{max}}^{r}n_{h}\sigma_{\nu} dr.
\end{equation}

Clearly, Equation (17) shows that the optical depth is function of radii,
but is not relative with $\theta$. Thus, it cannot correctly depict the penetration of the EUV
radiation from the face of day-side to night-side if the planets are
tidally locked by their host stars.
To calculate an accurate optical depth in the case of tidal locking,
two-dimensional radiative transfer calculations are needed. Here we present a simplified
treatment in which both the impact factor $q(r,\theta)=r\sin\theta$
and latitude $\theta$ determine the final optical depths. For each
grid point, the nodal points at every radii are found along the
opposite direction of stellar radiation, and the values of
latitude at those radii are determined (see Figure 1).
For a given grid point P(i, j), the nodal point at
the next radius $r_{i+1}$ can be expressed as P(i+1, j')
(Note that the impact factor $q(r,\theta)$ of the given grid point P(i, j) is same with
that of P(i+1, j').).
Generally, the latitude of P(i+1, j') is between grid points $\theta _{j}$ and
$\theta _{j-1}$ rather than consistent with the grid points of
angular direction. Finally, the corresponding number density of
P(i+1, j') at radius $r_{i+1}$  can be approximated as
$\overline{n_{h}}=\frac{n_{i+1,j}+n_{i+1,j-1}}{2}$.
Therefore, the optical depth along the z-direction (the direction of stellar radiation) can be approximated to

\begin{equation}
\tau^{i,j}_\nu=\int_{z_{max}}^{p}\overline{n_{h}}\sigma_\nu dz.
\end{equation}
The optical depth is set to $10^{10}$ when the impact factor $\emph{q}$ is smaller than $R_{p}$.
Equation (18) is used to calculate the radiative transfer of tidal locking.

\begin{figure*}
\epsscale{.60} \plotone{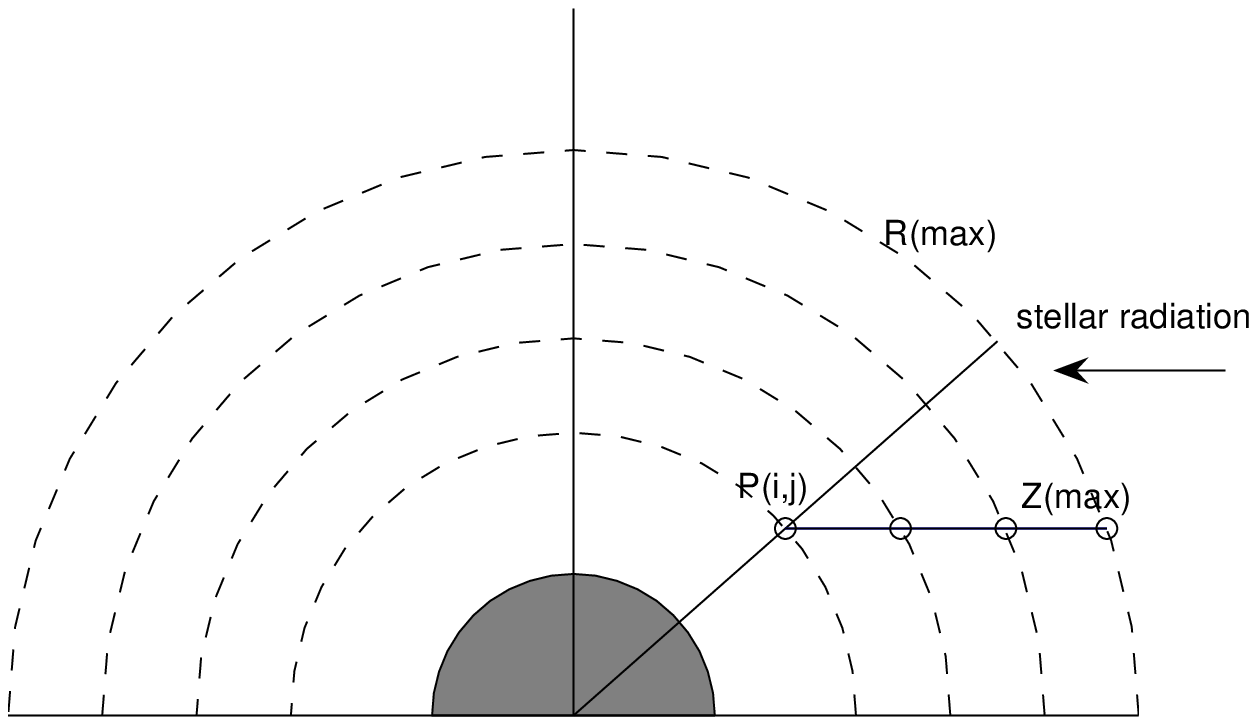} \caption{Schematic diagram of the
calculation method to the optical depth. The circles represent the
nodal points at other radii. $\emph{i}$ represents the radial
grid. $\emph{j}$ denotes the angular grid.}
\end{figure*}
The simplified calculations represent the
variations of the optical depth at all grid points. We believe that
small differences due to the approximation do not affect the results quantitatively.

\subsection{Numerical Method and Boundary Conditions}
The two-dimensional time-dependent multi-fluid model is integrated in time by
using the fourth-stage Runge-Kutta method. For the space derivative, central
difference scheme is used. To solve Equation (4), we must handle
those source terms in the continuity and energy equations
because of their stiffness. An explicit method could result in a
long time-marching and incorrect results. Thus, we choice a
semi-implicit method to deal with $S_{2}$ of Equation (4).

Equation (4) is discretized on a grid of 221 radial and 121 angular
cells. We selected exponentially spaced grid in radii,
\begin{equation}
r_{i}=qr_{i-1},
q=\textmd{exp}(\frac{\textmd{ln}\frac{\textmd{r}_{\textmd{IMAX}}}{r_{1}}}{\textmd{IMAX}-1})
\end{equation}
where $i$ is a number of a given grid point, IMAX is the maximum number of grid points,
and $r_{1}$ and $r_{IMAX}$ denote the first and last grid points.
The grid size increases exponentially with the increase of
radius. We used a uniform grid in $\hat{\theta}$ direction.

The lower boundary conditions at $r=R_{p}$ are fixed ($R_{p}$ is the planetary radius).
The effective temperatures of hot Jupiter estimated by Burrows et
al. (2001) are in the range of 1000 - 1500K, which is consistent
with the IR thermal emission measurements (Deming et al. 2005;
Charbonneau et al. 2005; Deming et al. 2006). Thus, we set
$T=1500$K at the bottom of the flow. Here the pressure at the base
of the wind is maintained on the level of $P_{R_{p}}=1$ dyn cm$^{-2}$.
The number density of H$^{+}$ at the lower boundary is set to
$n_{p}=10^{-4}n_{h}$. The calculation results are insensitive to the ratio of $n_{p}/n_{h}$.
Finally, we assumed $u_{\theta}=0$ at the bottom of the atmosphere.

We selected $r_{IMAX}=7R_{p}$ as the upper boundary, which approximately
denotes the stand-off distance of the planetary wind due to the interaction
with the stellar wind. At the
upper boundary, we used the outflow boundary conditions. For supersonic outflow, the
boundary condition is exact (Stone \& Norman 1992). In this paper, we only consider
the case of supersonic outflow. The planetary winds can be depressed to subsonic if the
stellar winds are strong enough (Garc\'{\i}a Mu\~{n}oz 2007; Murray-Clay et al. 2009).
A full description to the process is beyond the scope of
this paper. We will discuss the effect of the interaction with the stellar wind in
further work.

We are interested in steady-state solutions. For all hydrodynamic
solutions, the steady-state is reached when the relative change in
the conservative variables from one time level to the next drops
below $10^{-8}$. We use a normalized measure defined by (T\'{o}th
et al. 1998)
\begin{equation}
\bigtriangleup_{2}Q=\sqrt{\frac{1}{N_{\mathrm{var}}}\sum_{i=1}^{N_{\mathrm{var}}}\frac{\sum_{\mathrm{grid}}(Q_{i}^{n+1}-Q_{i}^{n})^{2}}{\sum_{\mathrm{grid}}(Q_{i})^{2}}},
\end{equation}
where $N_{\mathrm{var}}$ is the number of conserved variables $Q_{i}$, and the
superscripts indicate the number of time level.

\section{TEST THE ONE-DIMENSIONAL MODEL}

To test our model, we started from 1D planetary winds and applied
the 1D model to a typical and particular planet sample:
HD 209458b. We take the the radius  $R_{p}=1.4 R_{J}$ and the mass
of $M_{p}=0.7 M_{J}$, where $M_{J}$ and $R_{J}$ are the mass and radius
of the Jupiter. The semi-major axis $a$ is set to 0.047AU.
The mass of host star is $M_{*}=1.148M_{\odot}$ (http://exoplanet.eu/),
where $M_{\odot}$ is the mass of the sun.
Note that the tidal acceleration in the 1D model is written as
\begin{equation}
a_{r}=\frac{GM_{\ast}}{(D-r)^{2}}-G\frac{M_{\ast}}{D^{3}}(D-r),
\end{equation}
which is same as that of Garc\'{\i}a Mu\~{n}oz (2007).

Yelle (2004), Tian et al. (2005),
Garc\'{\i}a Mu\~{n}oz (2007) and Penz (2008) have developed
time-dependent one-dimensional hydrodynamic models. Thus, a direct
comparison with their results can validate the reliability
of our model.

With the stellar and planetary physical parameters,
the time-dependent results are shown in Figure 2. It is clear from
Figure 2 that the time-dependent model predicts the same trends
of temperature, velocity and particle number density as do the other
models. Those studies that the temperature can attain
8000-10000K at 1.5$R_{p}$ are in agreement with our results.
The velocities predicted by our model are higher than those of
Yelle (2004), but are similar with those of Tian et al. (2005),
Garc\'{\i}a Mu\~{n}oz (2007) and Penz (2008).
The fact that as much as 50\% of hydrogen at $r=2.R _{p}$ is ionized is a
consequence of being irradiated by the host star. Due to lack of photochemistry,
the ionization state cannot be predicted accurately by our model. In fact, the
atmosphere can be thought of as if made up solely of hydrogen if
the amounts of heavy constituents are small enough.

The total mass loss rate
of 3.2$\times$ 10$^{10}$ g s$^{-1}$ predicted by our model is comparable with the those of Yelle(2004), Tian et al. (2005),
Penz (2008) and Garc\'{\i}a Mu\~{n}oz (2007) (In Table 1, we summarized the cases with P$_{R_{p}}$=10 dyn cm$^{-2}$ and $P_{10.7}=200$.
The mass loss rate of 2.2$\times$ 10$^{11}$ g s$^{-1}$ is comparable with that of Garc\'{\i}a Mu\~{n}oz (2007)).
Some chemical processes are not included in our model. Thus,
the net chemical expense of energy cannot be calculated. However, the escape of particles
in the atmosphere of planet should be driven mainly by irradiation of host star.
Our results show also that the mass loss rates are in good agreement with those
of others, thus the effect of chemical energy
could be slight.

Yelle(2004) and Garc\'{\i}a Mu\~{n}oz (2007) have found that the escape rate is energy limited, namely, the
escape rate varies roughly with the stellar flux. To test the effect of stellar
flux, we multiplied
the solar spectrum by factors of 2.2 and recalculated the mass loss rate. We found that the mass loss rate predicted by the model of 2.2 times solar spectrum
is a factor of 3.7 higher than that of solar spectrum. Similar conclusions have also been found by Penz et al. (2008). From Table. 1 in Penz et al. (2008),
it is clear that the ratio of mass loss rates are not proportional to the stellar flux. The difference can be explained by a few reasons.
First, photochemistry has been neglected in our model and Penz et al. (2008). However, it is unclear whether the photochemistry can affect
the ratio of mass loss rates. We will discuss this in future work.
Second, we note that the net heating efficiency in our model is defined as $\eta_{\nu}=(h\nu-13.6)/h\nu$, which
varies with the frequency, but not with radii. In fact, the heating efficiencies should be varied with radii (Yelle 2004).
However, the Equation (11) of Garc\'{\i}a Mu\~{n}oz (2007) expresses that they assumed a net heating efficiency of 100\% for all single photons.
Our calculations show also that the ratio of mass loss rates decrease to 2.58, roughly equal with the ratio of incident stellar flux,
if the net heating efficiency is assumed to 100\%. This hints that the heating
efficiency plays an important role in simulating the process of the particle escape. Thus, the realistic heating
efficiencies of photons in the atmosphere of extra-solar planets should be investigated further by
detailed radiative transfer calculations.

\begin{figure*}
\epsscale{.80} \plotone{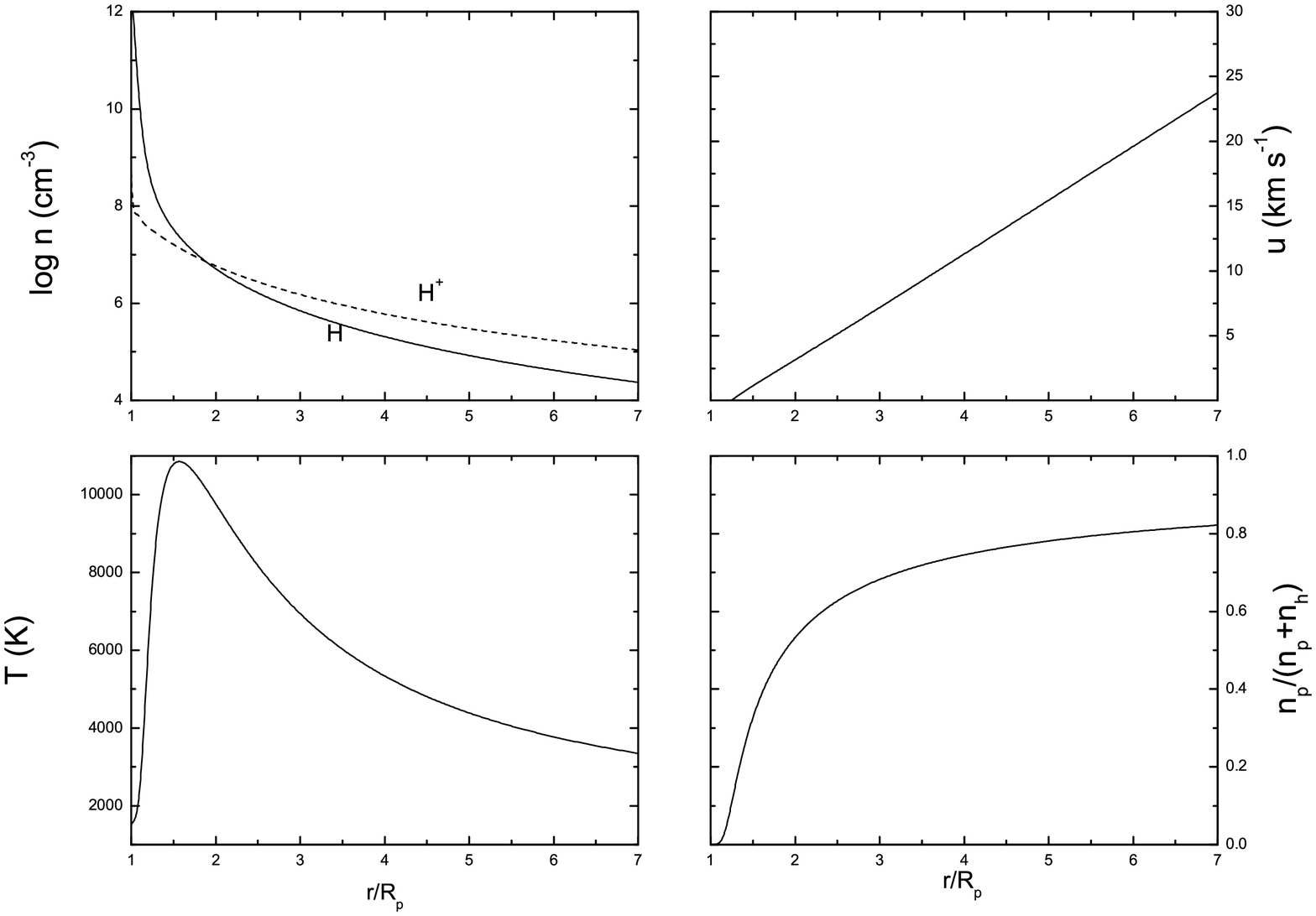}  \caption{The one-dimensional wind model for HD
209458b. Number densities
(upper left), velocities (upper right), temperatures (lower left)
and the ionization fraction (lower right) are plotted as the functions
of altitude.}
\end{figure*}

An interesting issue is whether a single photon of 20 ev as done in the models of Murray-Clay et al.
(2009) and Paper I can denote the stellar EUV irradiation. To test this assumption, we recalculated the model with a single
photon of 20 ev (here we assumed that the EUV flux is the same, but we only calculated the radiative transfer of photon of 20 ev) 
and found that the mass loss rate is 6.1$\times$ 10$^{10}$ g s$^{-1}$ which is a factor 2 greater
than the value predicted by using whole stellar EUV flux. Moreover, our calculations found that the maximum
mass loss rate (9.55$\times$ 10$^{10}$ g s$^{-1}$) is produced when the photon energy
is about 26 ev, but photons with higher or lower energy will result in lower mass loss rates.
Murray-Clay et al. (2009) predicted a mass loss rate of 3.3$\times$ 10$^{10}$ g s$^{-1}$ when incident
flux is decreased to F$_{UV}$=450 erg s$^{-1}$ cm$^{-2}$ integrated on the the solar flux between
photon energies of 13.6 ev to 40 ev). Applying the similar stellar flux (F$_{UV}$=540 erg s$^{-1}$ cm$^{-2}$, the value
is half of the whole incident flux ), we found that the results calculated by the whole EUV flux
can be matched well with a single photon of 22ev.
By comparing the ionization, the location of sonic point, temperature structure and mass loss rates,
we found that the effect of the whole EUV irradiation can not be substituted by a single photon unless
the EUV flux is decreased a factor of 2.

Thus, we can summarize that the effect of the whole EUV irradiation cannot be
represented simply by a single photon because
different photons penetrate to different depth and supply different
heating. The approximation of a single photon erases the characteristics of different photons.
In order to obtain reliable results, the contributions from all photons should be included fully.

\section{THE TWO-DIMENSIONAL MODELS}

\subsection{The effect of tidal force}
In this section we solved the two-dimensional hydrodynamic equations.
To separately discuss the effect of the tidal force,
we assumed the case of non-tidal locking, thus the one-dimensional radiative transfer is used (Equation (17)).
However, the opacities on the day and night
sides may be very different. We will further discuss it in Section 4.2.

The left panels of Figures 3 and 4 show the images of the density (with velocity vectors)
and temperature. Note that the density is the sum of hydrogen atoms and ions.
In the 2D model, the number density of hydrogen decreases but the ionization fraction increases with
radius as does one-dimensional hydrodynamic model.
The temperature rises with an increase in
radius up to r/R$_{p}=2$. The two-dimensional
models show that wind temperature can attain
10000K at about 1.5$R_{p}$ and then drop to 3000K.

\begin{figure}
\begin{minipage}[t]{0.5\linewidth}
\centering
\includegraphics[width=2.9in,height=2.2in]{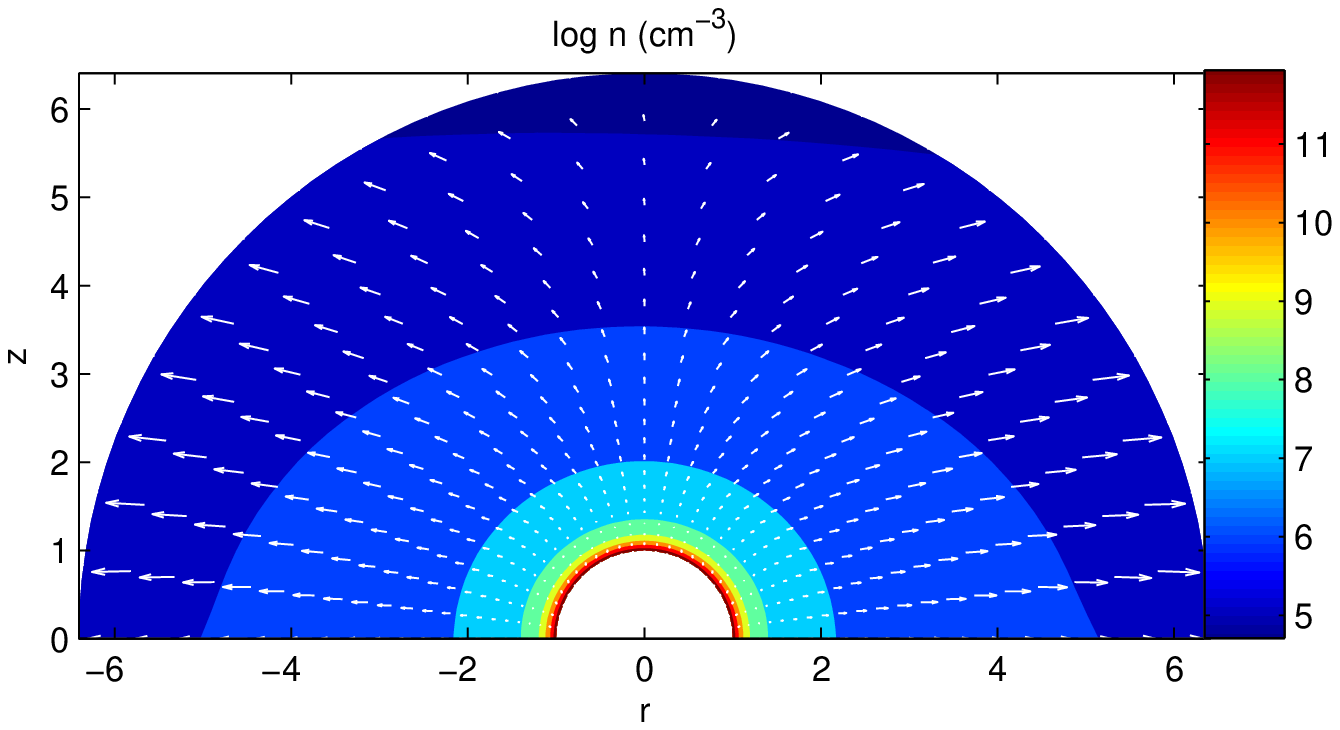}
\end{minipage}
\begin{minipage}[t]{0.5\linewidth}
\centering
\includegraphics[width=2.9in,height=2.2in]{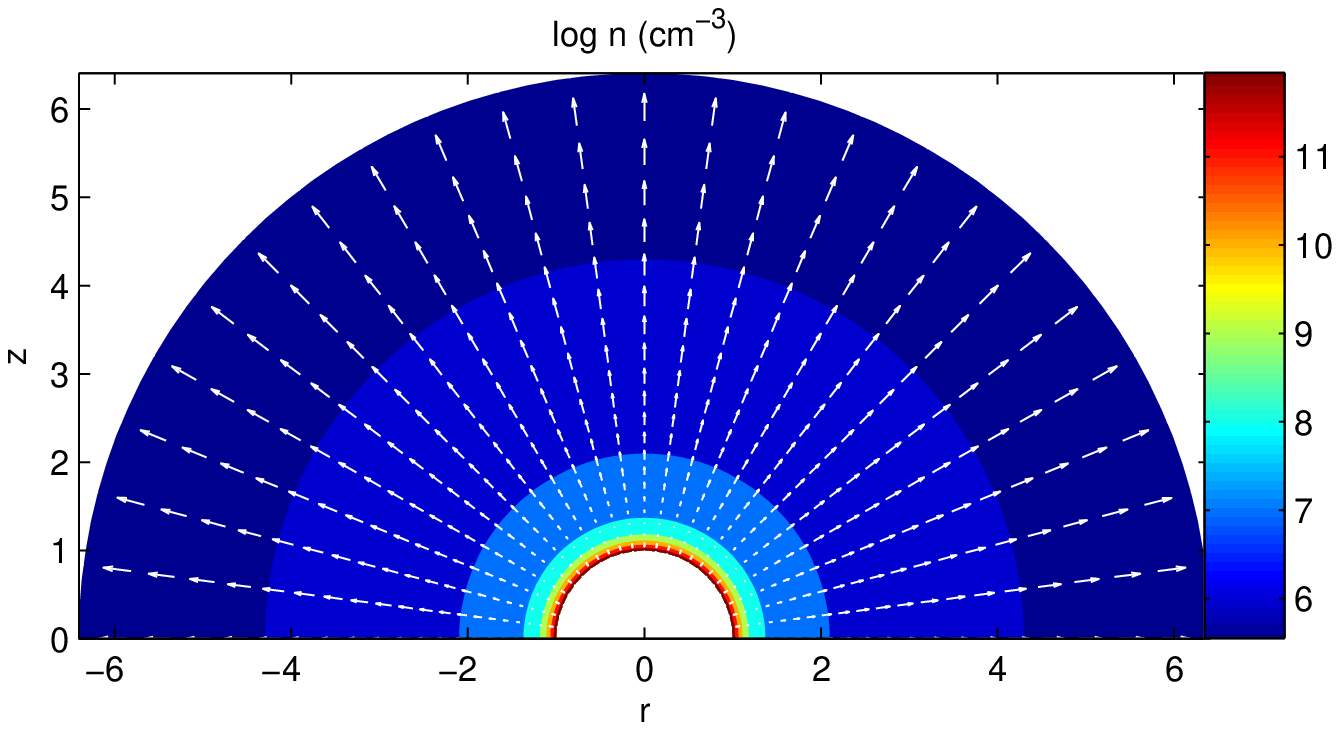}
\end{minipage}
\caption{The results of two-dimensional hydrodynamic model with
one-dimensional radiative transfer. Left: density and velocity
vectors for the case with tidal forces. Right: density and velocity
vectors for the case without tidal forces.
The host star is located toward the right
(corresponding to $\theta$ = 0).}
\end{figure}

\begin{figure}
\begin{minipage}[t]{0.5\linewidth}
\centering
\includegraphics[width=2.9in,height=2.2in]{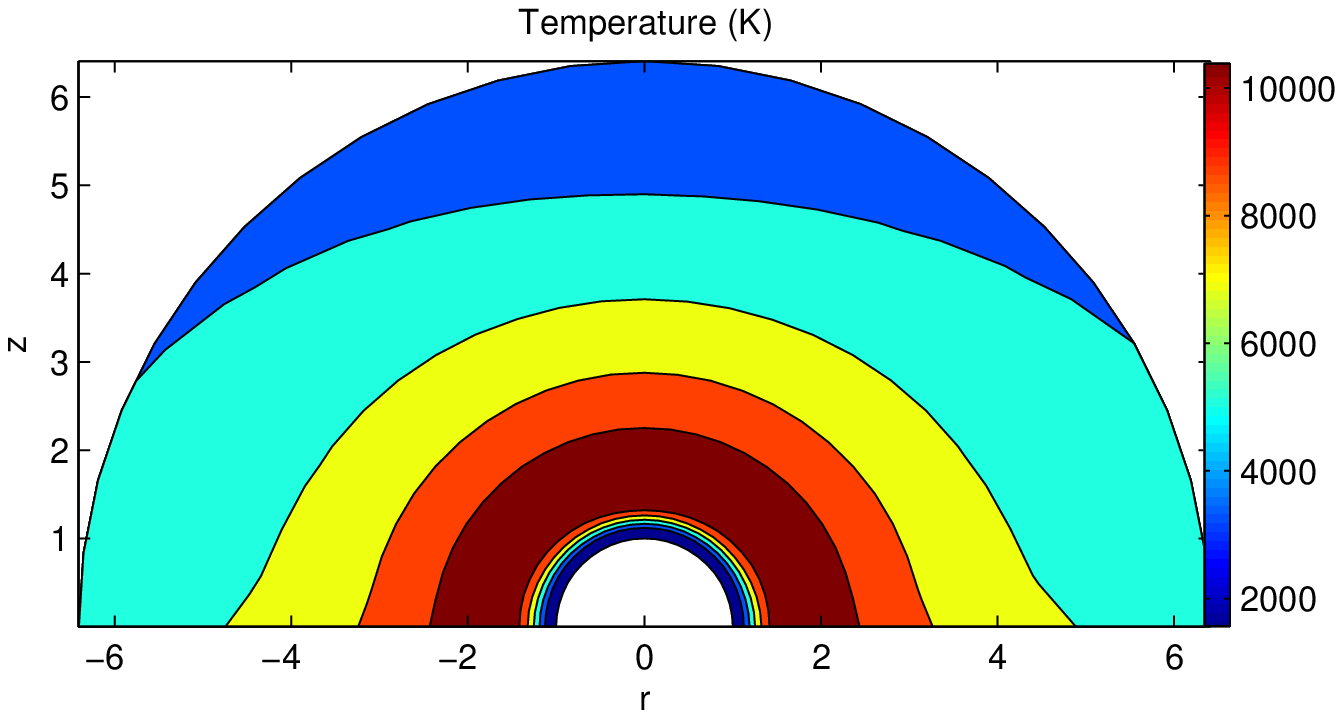}
\end{minipage}
\begin{minipage}[t]{0.5\linewidth}
\centering
\includegraphics[width=2.9in,height=2.2in]{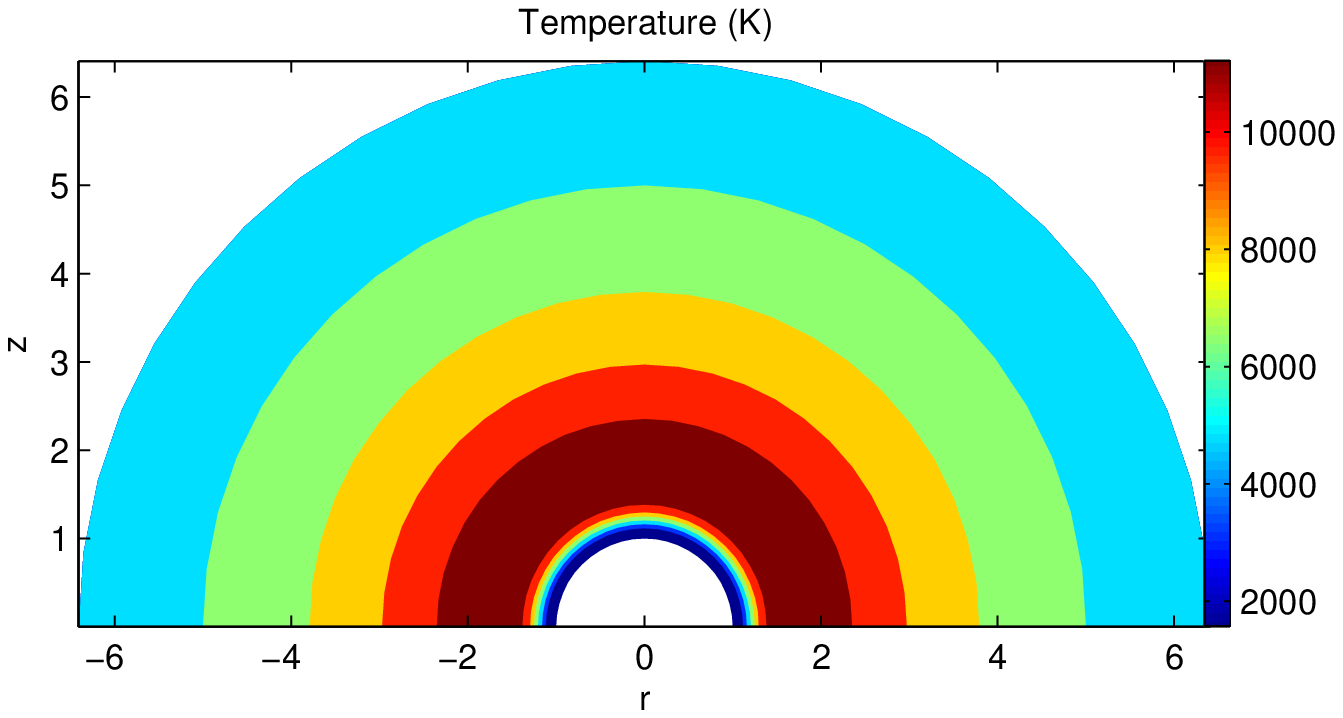}
\end{minipage}
\caption{The results of two-dimensional hydrodynamic model with
one-dimensional radiative transfer. Left: temperature distributions for the case with tidal forces.
Right: temperature distributions for the case without tidal forces.
The host star is located toward the right
(corresponding to $\theta$ = 0).}
\end{figure}


The mass loss rates are defined as
\begin{equation}
\dot{M}=2\pi m_{\mathrm{H}}(r_{max})^{2}[\int_{\pi}^{0} n_{h}u_{{r}}
\sin\theta d\theta+\int_{\pi}^{0} n_{p}u_{r}
\sin\theta d\theta],
\end{equation}
where $r_{\mathrm{max}}$ is the upper boundary. We obtained a mass
loss rate of $\dot{M}=1.72\times 10^{10}$ g s$^{-1}$ which is a factor
of 2 smaller than that calculated by one-dimensional hydrodynamics
model.

From Figures 3 and 4, a prominent characteristic is that the contours of the density and temperature
are nearly axis-symmetric. It is clear from the left panel of Figure 3 that the pattern of flows are non-radial.
The flow vectors show curve toward the equator, and the radial velocities
and densities are very low near the polar regions. We have demonstrated this in Figure 5
in which $\frac{n({\theta})}{n({\theta=0})}$ and $\frac{u_{r}({\theta})}{u_{r}({\theta=0})}$ at r=7$R_{p}$ are showed
as a function of $\theta$. It is clear that the density and radial velocity
decrease rapidly with an increase in $\theta$. The values of $\frac{n({\theta})}{n({\theta=0})}=0.04$ and
$\frac{u_{r}({\theta})}{u_{r}({\theta=0})}=0.2$ at $\theta$=90$^{0}$ reflect the fact
that most particles escape the planet from the zones of low latitude.
In fact, the zones of middle-high latitude(45$^{0}$-135$^{0}$)
only contribute to 20\% of the mass loss rate due to their low density and radial velocity.
As a consequence, the mass loss rate
of the two-dimensional hydrodynamics is only half of that of one-dimensional hydrodynamics.
Figure 6 further shows the effect of tidal forces at different radii.
Due to strong gas pressure, the radial velocities are evidently higher than
the meridional velocities at the base of the wind (r/R$_{P}$=2). However, the ratios of $u_{\theta}/u_{r}$
rapidly increase with an increase in radius. At r/R$_{P}$=4, the ratio can attain 0.6 at $\theta$=70$^{\circ}$,
and the value even approaches 1 at r/R$_{P}$=7. The trend shows how the tidal forces enforce the flows moving
in horizonal direction when the gas pressures decrease with an increase in radius.
In addition, the peaks of $u_{\theta}/u_{r}$ are almost emerge at $\theta$=70$^{\circ}$
rather than $\theta$=90$^{\circ}$ can be attributed to the balance of orbital motions and gravity accelerations of host star.

\begin{figure}
\epsscale{.60} \plotone{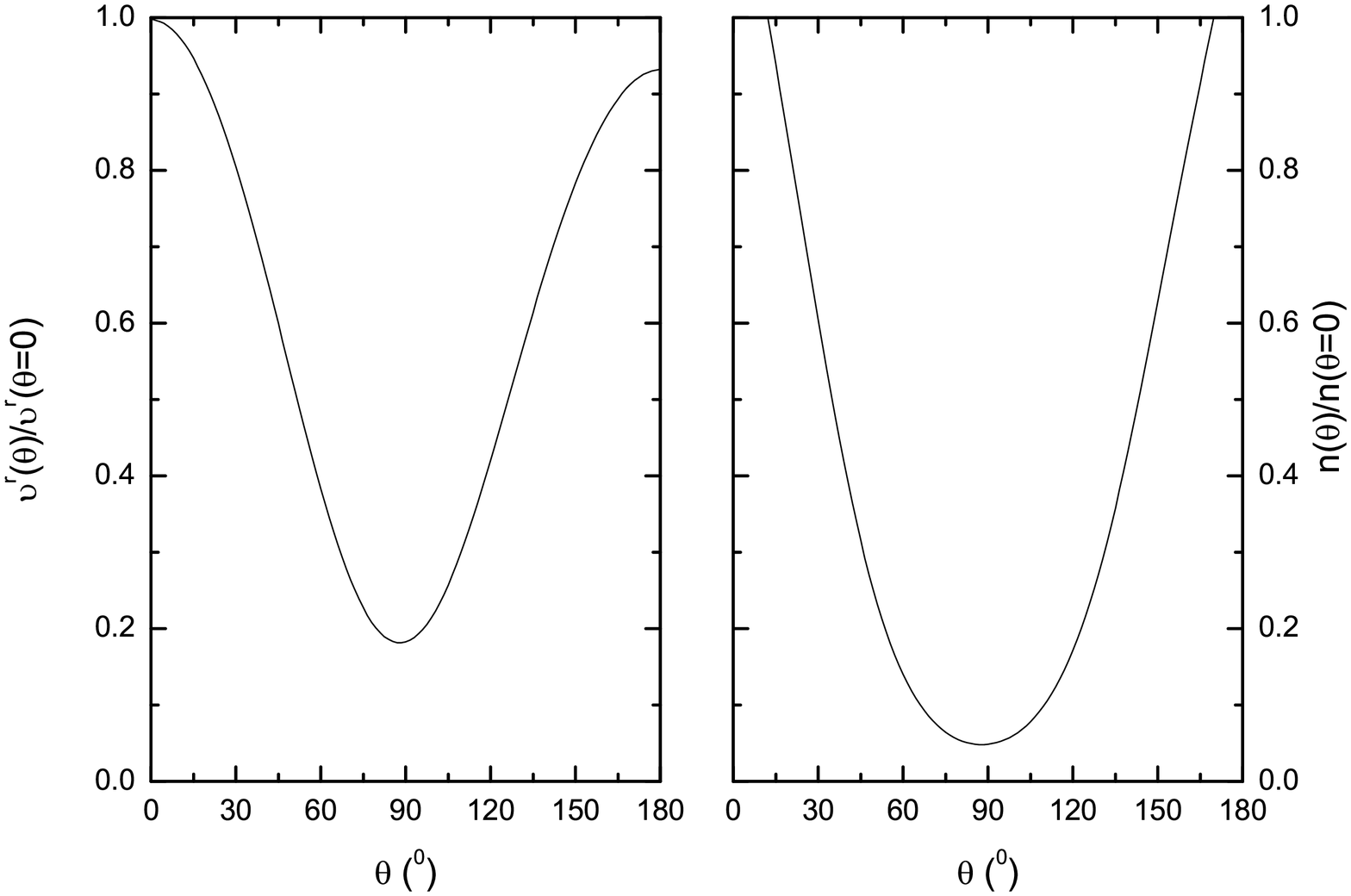}
\caption{Left: the radial velocity $\upsilon^{r}$ as a function of $\theta$ at r/R$_{p}$=7.
Right: the distributions of number density as a function of $\theta$ at r/R$_{p}$=7.}
\end{figure}

To further explore the effect of tidal force,
we neglected the tidal force and recalculated the model.
The right panels of Figures 3 and 4 show spherically symmetric distributions
for density, velocity and temperature. Comparing the left and right panels of Figures 3 and 4,
we can find that the tidal forces not only supply
strong accelerations in $\hat{\theta}$ direction but also decrease the radial velocity in the polar regions.
Therefore, we can draw the conclusion that the tidal forces not only impel the non-radial flows in the planetary atmosphere but also inhibit the
escape of particles from the polar regions
so that most particles escape the bound of the planet from the zones of low latitude.

\begin{figure}
\epsscale{.60} \plotone{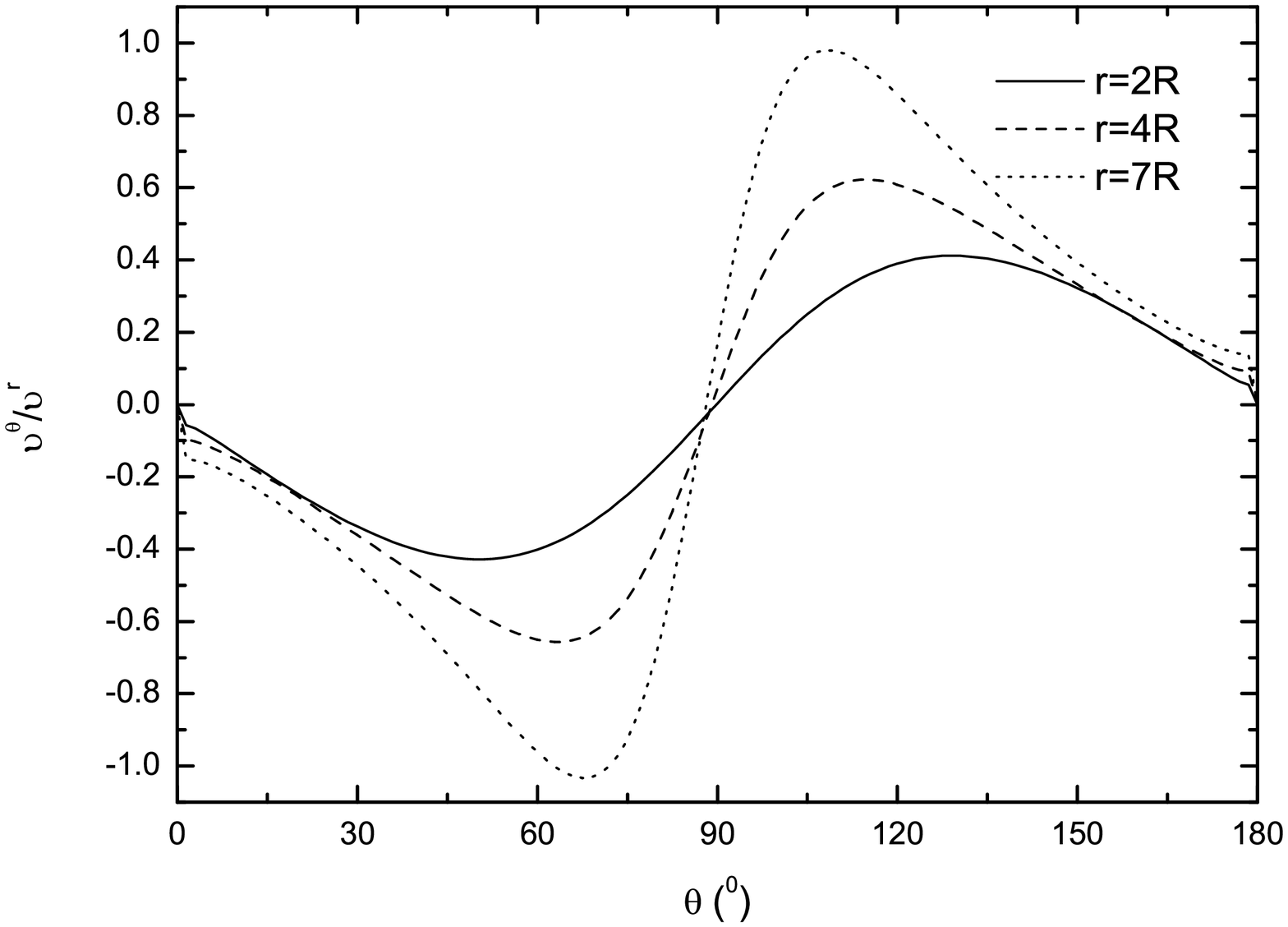}
\caption{The values of $u_{\theta}/u_{r}$ as a function of $\theta$ at r/R$_{p}$=2, 4 and 7.}
\end{figure}

\subsection{The Case of Tidal Locking}

In Section 4.1, we discussed the case of non-tidal locking.
However, about 25\% extra-solar planets are very close to their host stars
thus they should be tidally locked. To inspect the case,
we recalculated the two-dimensional planetary winds by
using two-dimensional radiative transfer (Equation (18)),
which appropriately depicts the irradiation of the host star.

Figure 7 displays that the decline of density on the night sides with the
radius is slightly slower than that on day sides.
At large radii, the densities of polar regions are
lower than that of the regions of low-middle latitude.
Close to the planet, the
meridional velocities dominant the velocity field and transfer the mass
and energy from the day side to the night side (see the left panel of Figure 7 for details).
The appearance of strong
non-radial flows at the base of the wind can be
explained by the variations of the optical depth.  The optical depth of single photon of 20 ev
for two-dimensional radiative transfer is shown
in right panel of Figure 8. The contours of
the optical depth extend toward a larger radius from the day side to
the night side. When the impact factor $\emph{q}$ approaches the
radius of the planet, the optical depth is in the order of magnitude
of $10^{-1}-10^{1}$. The heating is proportional to
e$^{-\tau}$ (Equation (15)), therefore,  more energy can be received on
the day side. With an increase in
radius, the radial velocities increase gradually.
Finally, the pattern of flows at the base of the wind is nearly
in the direction of meridian. In the upper atmosphere, the non-radial
flows are comparable with the radial flows,
but the radial velocities are stronger in the zones of low-latitude.
\begin{figure}
\begin{minipage}[t]{0.5\linewidth}
\centering
\includegraphics[width=2.9in,height=2.2in]{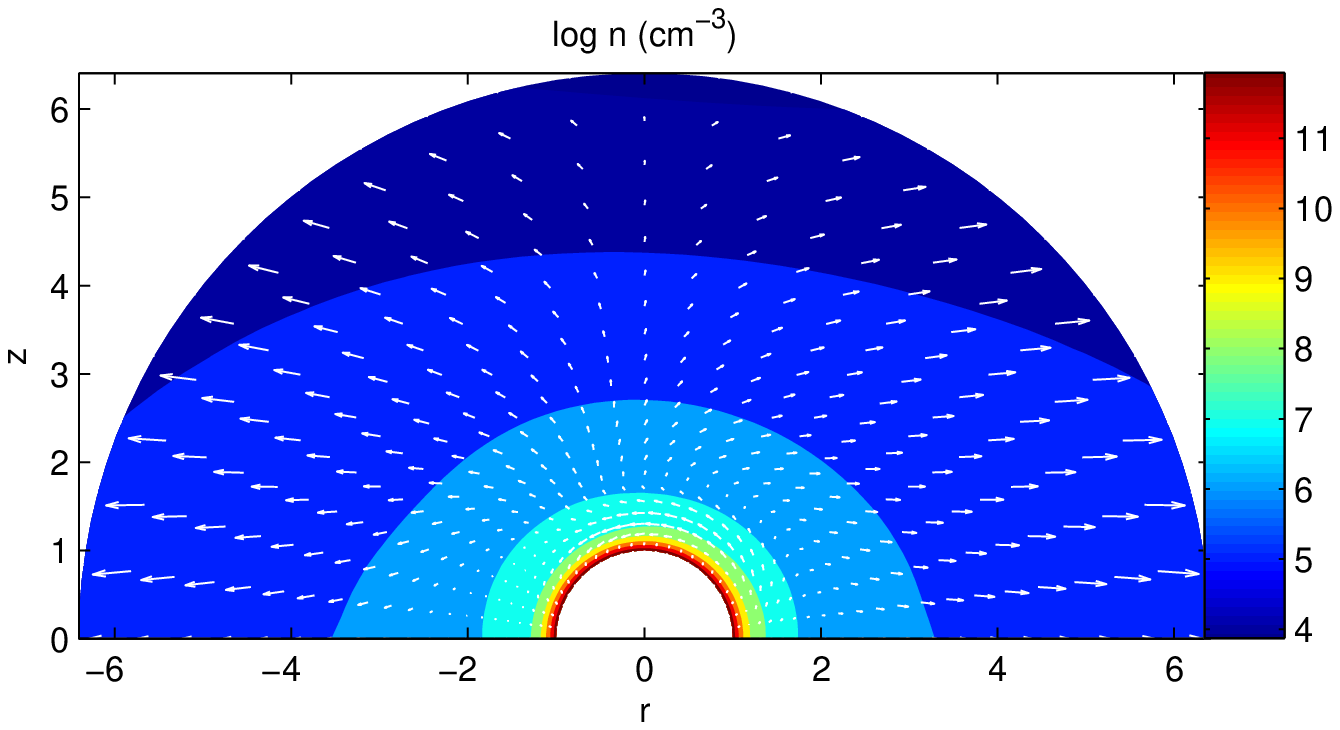}
\end{minipage}
\begin{minipage}[t]{0.5\linewidth}
\centering
\includegraphics[width=2.9in,height=2.2in]{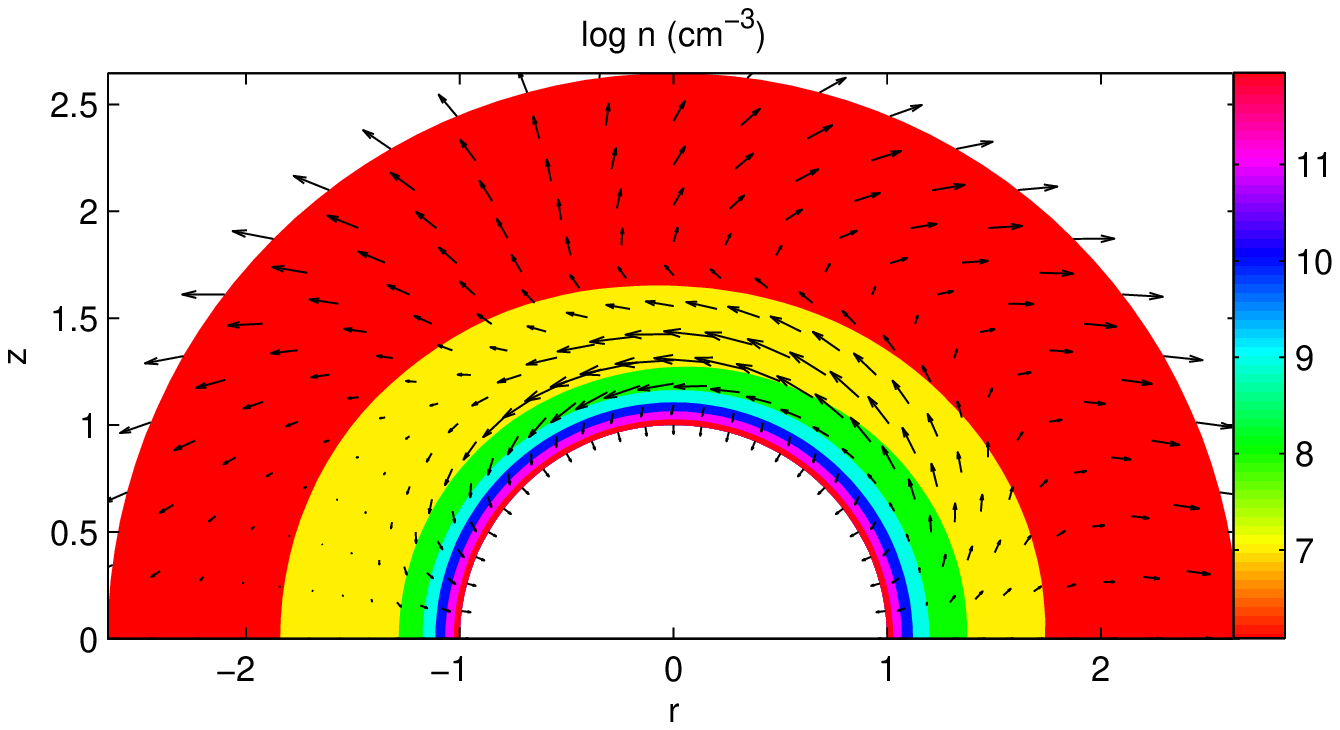}
\end{minipage}
\caption{The results of the two-dimensional hydrodynamic model
with two-dimensional radiative transfer.
Left: density and velocity vectors at 1 $\leqslant$ r/R$_{p}$ $\leqslant$ 7.
Right: density and velocity vectors at the atmospheric base (1 $\leqslant$ r/R$_{p}$ $\leqslant$ 2.6).}
\end{figure}

\begin{figure}
\begin{minipage}[t]{0.5\linewidth}
\centering
\includegraphics[width=2.9in,height=2.2in]{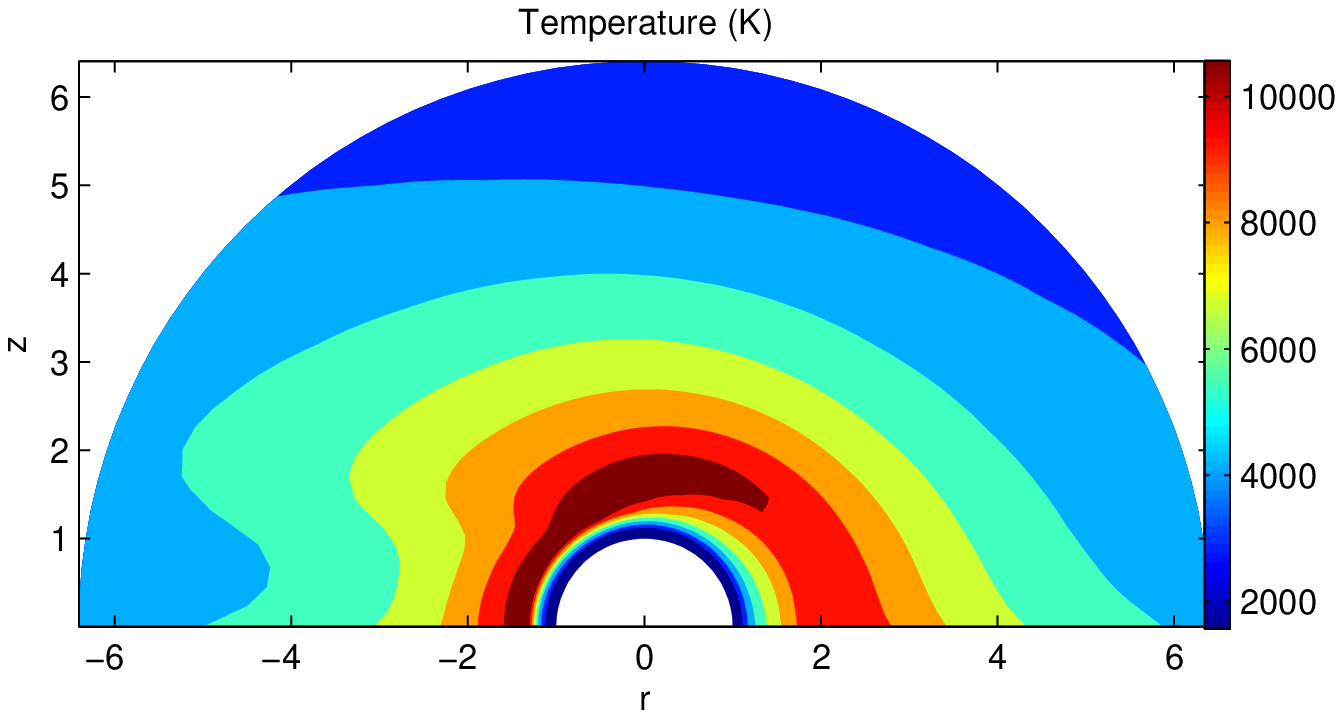}
\end{minipage}
\begin{minipage}[t]{0.5\linewidth}
\centering
\includegraphics[width=2.9in,height=2.2in]{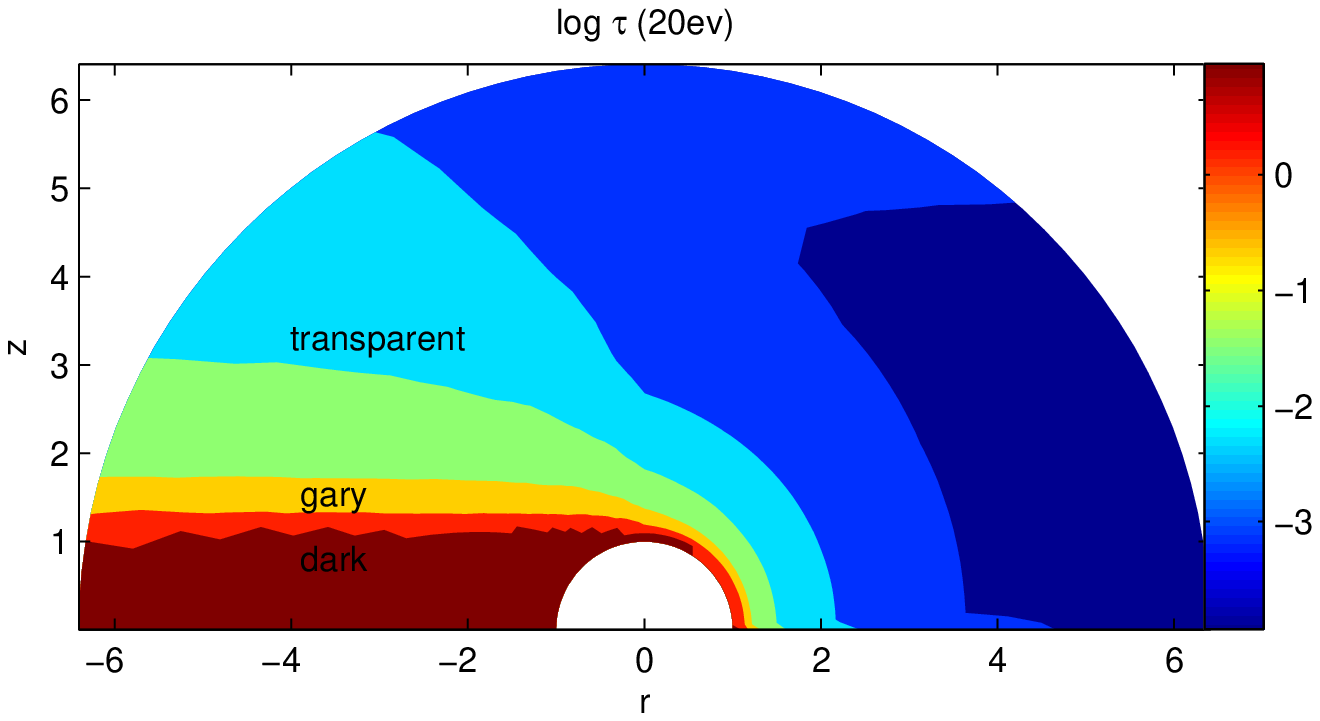}
\end{minipage}
\caption{The results of the two-dimensional hydrodynamic model
with two-dimensional radiative transfer. Left: the
distributions of temperature. Right: the distributions of the optical depth.
In the right panel the optical depth lager than
10 is marked with "dark". The "gray" region denotes the case of
10$^{-1}<\tau <$ 1. In the transparent region, the optical depth
is smaller than 10$^{-1}$.  }
\end{figure}

The distributions of temperature are also non-spherically symmetric due
to the influence of the optical depth (the left panel of Figure 8).
Evidently, the distributions of temperature in the night side are
controlled by the optical depth. The radiation of the star cannot transmit
to those regions where the optical depths are greater than unity. Thus
those regions of lower temperature are located at the proximity of impact
factor q$\sim$ 1 (lower left of right panel of Figure 8).
The hottest regions is located at $60^{\circ}\lesssim \theta
\lesssim180^{\circ}$ at r/R$_{p}=1.3-1.8$, which is
caused by the balance of PdV work (P$\nabla\cdot \mathbf{v}$), heating, and cooling.
To explain the variations of temperature, the contributions of energy at r/R$_{p}$=1.35 by PdV work, heating,
cooling and work $W_{ext}$ done by external forces (gravity of the planet and tidal forces) are shown in Figure 9. Heating from
the irradiation and PdV work continuously decline to zero from the day side to night side.
However, the work done by external forces continuously rises
with the increase of $\theta$.
Due to $de=dQ-PdV+W_{ext}$, this hints that external forces can do positive work
due to the contraction of gas, which results in the increase of gas temperature
(In generally, PdV work expands the atmosphere and lowers the gas temperature.).
Therefore, the cause of temperature increase from the day to night side can be explained by the increase of
net heating rate. While $105^{\circ}\lesssim \theta
\lesssim125^{\circ}$, the temperature slightly declines with the decrease of net heating rate.
And eventually, the temperature becomes a constant because the net heating is
maintained at a stable level.

\begin{figure*}
\epsscale{.80} \plotone{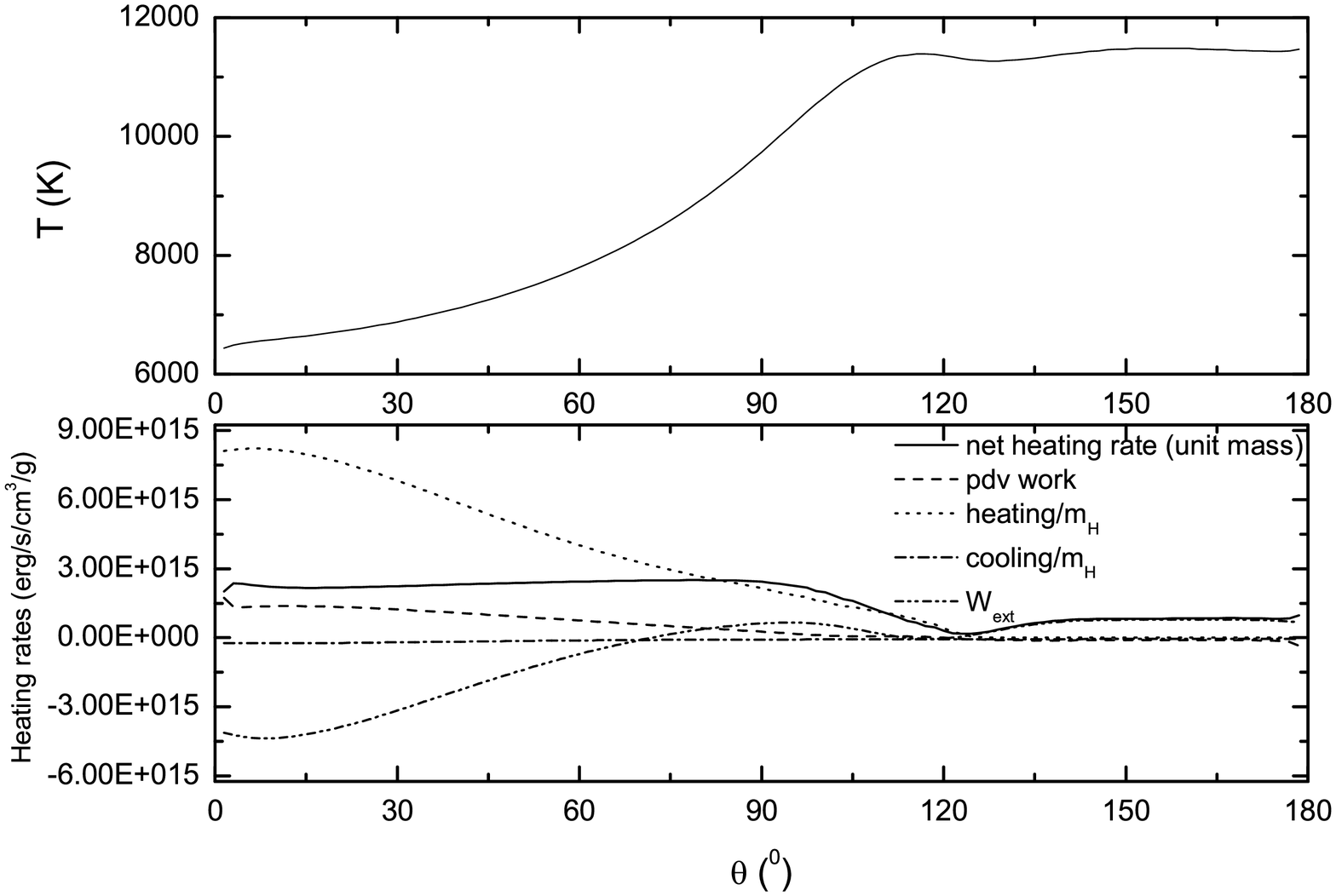}  \caption{Temperatures and heating rates (unit mass) in the model.
In the lower panel, the solid line represents the net heating rate in the atmosphere.
The dashed, dotted and dash-dotted lines denote the PdV work, heating from host star and
cooling. The term W$_{ext}$ (dash-dot-dotted) represents the work made by the gravity of planet and tidal forces.}
\end{figure*}

We also display the angular slices of the density and velocity
at the different radii in Figures 10-12. For comparison,
the winds of one-dimensional radiative transfer are also included
(solid lines). Seen from Figure 10, the difference in the number density between one- and
two-dimensional cases is clear. The total number
densities of the two-dimensional radiative transfer are much less than
those of one-dimensional radiative transfer, by a factor of 2-10 (depend on $\theta$).
Thus, the mass loss rate also decreases a factor of 4, only $4.3\times
10^{9}$ g s$^{-1}$.

The changes of radial velocity with the angles demonstrate different
features between the upper and lower atmospheres. In the upper
atmosphere (r/R$_{p}=4$ and 7), the radial velocities decline with
an increase in angle but rise at $\theta=90^{\circ}$ again (the left panel of Figure 11).
The minimum and maximum radial velocities at r/R$_{p}=7$ are 4 ($\theta\approx90^{\circ}$) and 21($\theta\approx0^{\circ}$) km
s$^{-1}$, respectively. In comparing with the case of one-dimensional radiative transfer,
the differences decrease with an increase of radius.

However, we also note that the radial velocities at r/R$_{p}=2$
almost decline to zero with an increase of $\theta$ in the case of two-dimensional radiative transfer.
Moreover, the complex behaviors at r/R$_{p}=$1.5 reflect the difference between two-
with one-dimension radiative transfer (the left panel of Figure 11). This can be attributed
to the effect of optical depth. At r/R$_{p}\approx1.5$ and 2, the optical depths can
vary few order of magnitude from the day side to the night side, thus the processes of radiative transfer
play an important role.

In Figure 12, interesting trends in the angular
velocity can be found. First, for both one- and two- dimensional
radiative transfers, the meridional velocities show the sine-like profile
in the middle-upper atmosphere (the left of Figure 12) and are always negative (positive)
if $\theta$ is smaller (greater) than $\pi/2$.
In contrast, the meridional velocities are always positive at small radii
for the case of two-dimensional radiative transfer (the right panel of Figure 12), which reflects the transfer
of mass and energy from the day side to the night side. Second, the
meridional velocities are comparable with (or even large than) the radial
velocities in the middle-high latitude regions of the upper atmosphere. Third,
the meridional velocities dominant the velocity field in the base of the wind for the
case of two-dimensional radiative transfer, but we have not found similar phenomena in the case of
of one-dimensional radiative transfer.

Therefore, from Figures 7-12, we can conclude that the two-dimensional radiative
transfer prominently affects the properties of flow. 
The influence of velocity field is remarkable at the base of the wind. Due to the large
optical depth on the night side, the density decrease a factor of few
so that the mass loss rate is a factor of 4 lower than that of one-dimensional radiative
transfer. 
However, the atmosphere becomes transparent with
an increase in the radius so that
there is moderate difference in the upper atmosphere between the one- and two-dimensional radiative transfer.

\begin{figure*}
\epsscale{.80} \plotone{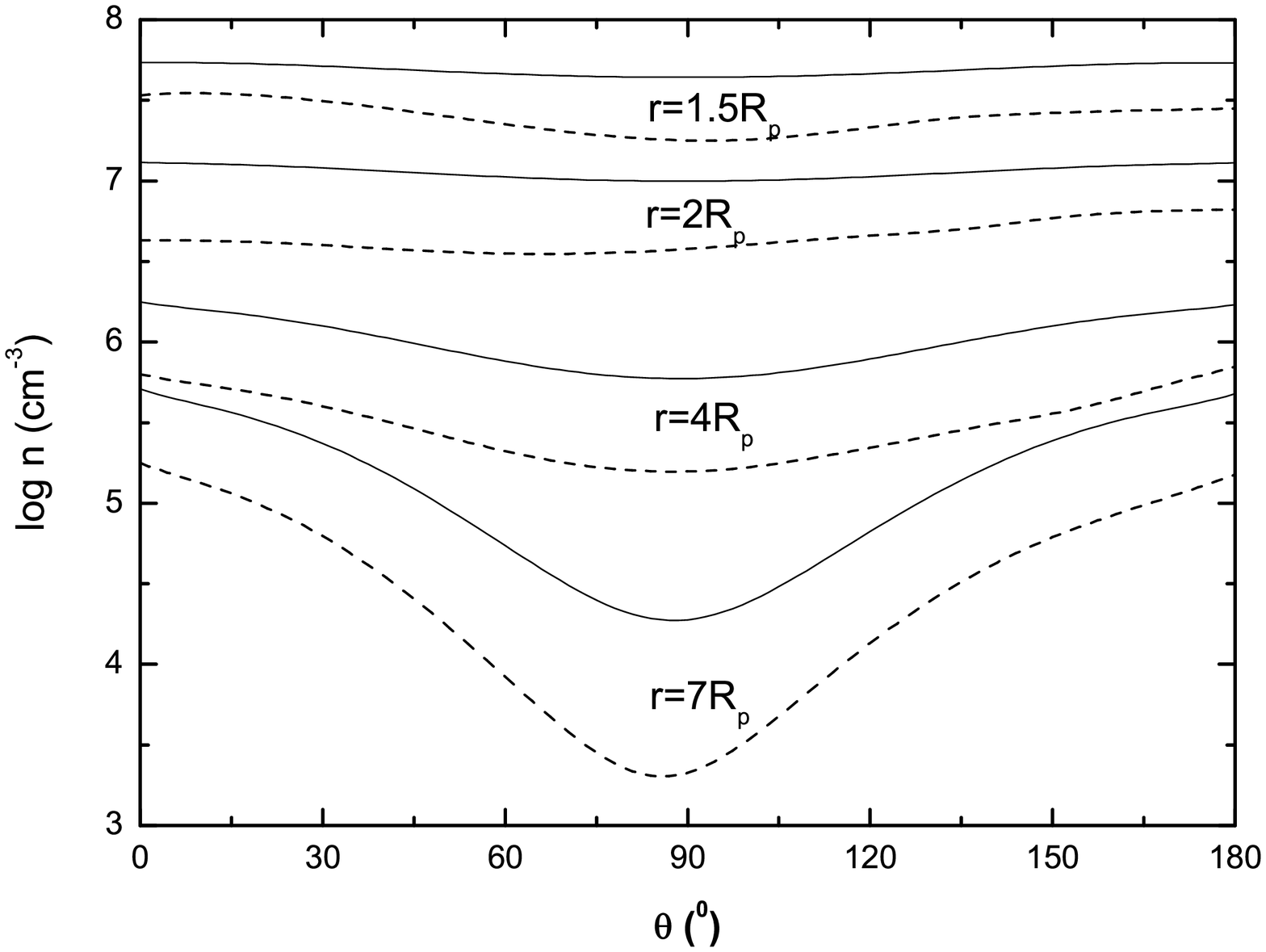}  \caption{ Number densities at
r/R$_{p}$ =1.5, 2, 4 and 7 (from top to bottom). Solid lines represent
the case of one-dimensional radiative transfer. Dashed lines represent
the case of two-dimensional radiative transfer. Note: The number densities are the
sum of hydrogen atoms and ions.}
\end{figure*}

\begin{figure*}
\epsscale{1.00} \plotone{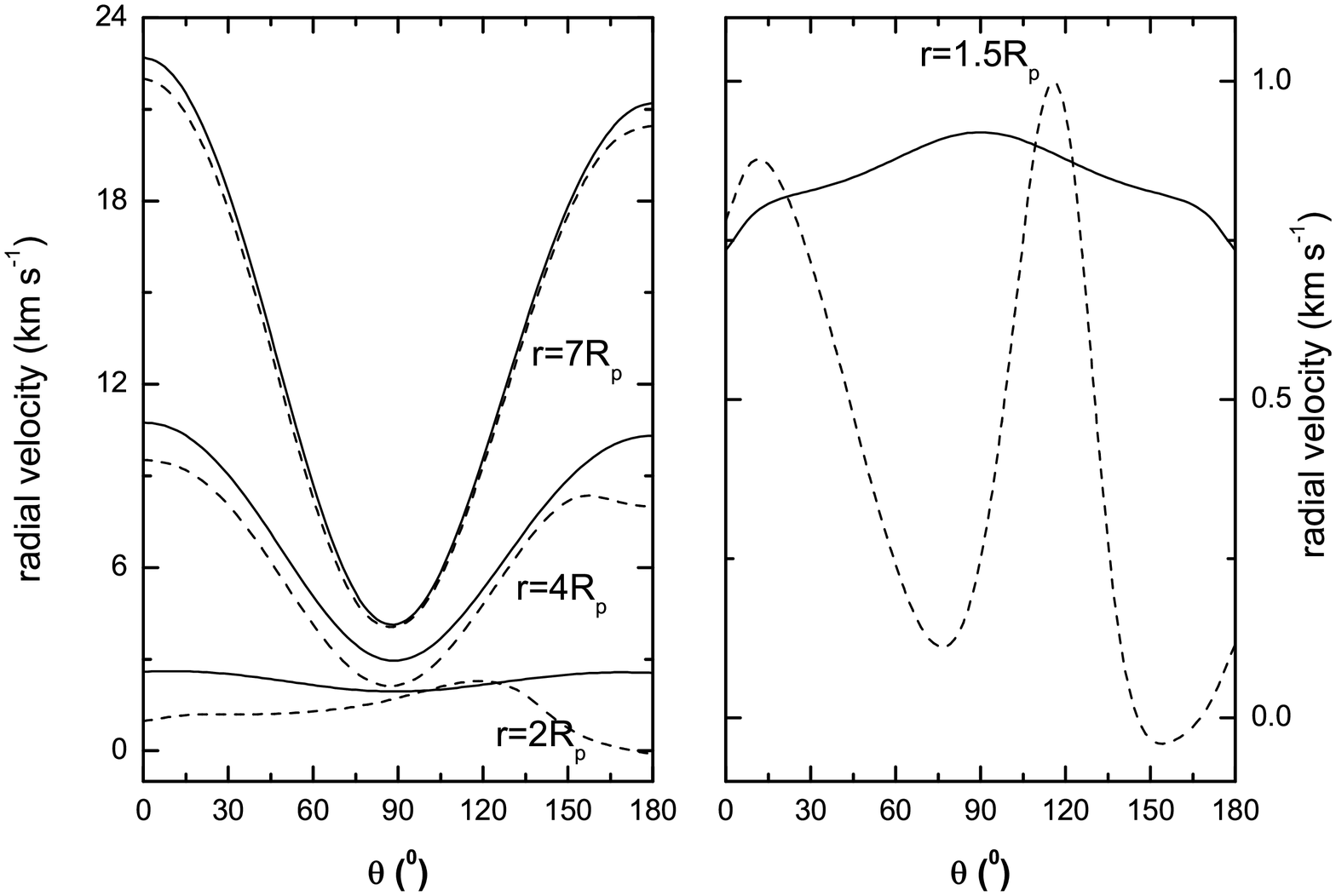}  \caption{Left: radial velocities at
r/R$_{p}$ =2, 4 and 7. Right: radial velocities at
r/R$_{p}$ =1.5. Solid lines represent
the case of one-dimensional radiative transfer. Dashed lines represent
the case of two-dimensional radiative transfer. }
\end{figure*}

\begin{figure*}
\epsscale{1.00} \plotone{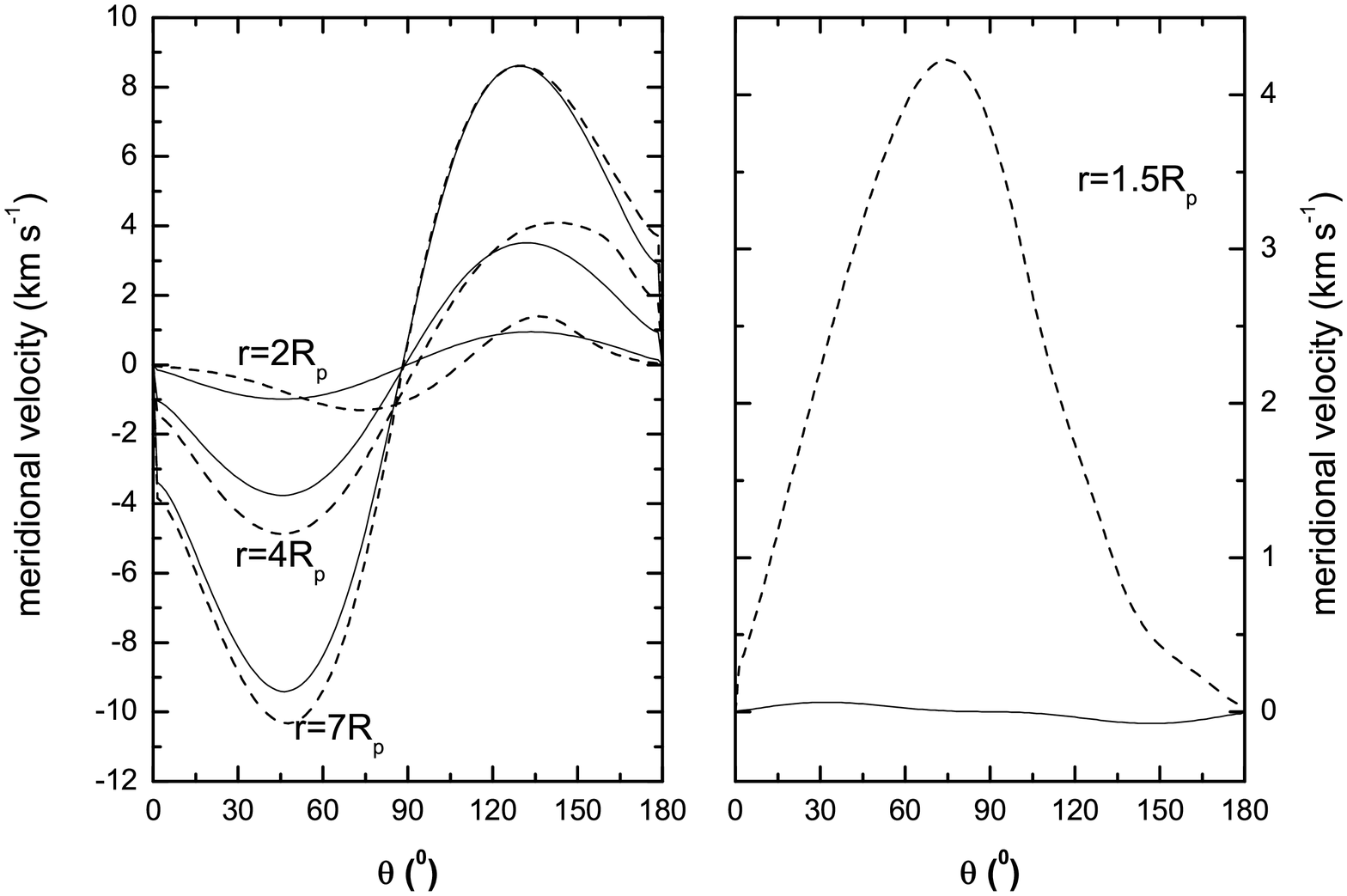}  \caption{Left: meridional velocities at
r/R$_{p}$ =2, 4 and 7. Right: meridional velocities at
r/R$_{p}$ =1.5. Solid lines represent
the case of one-dimensional radiative transfer. Dashed lines represent
the case of two-dimensional radiative transfer.}
\end{figure*}

\section{DISCUSSIONS}
We extended our calculations to different cases.
The mass loss rates predicted by our models are summarized in Table 1.

\subsection{Boundary conditions  }
So far, our models used P$_{R_{p}}$=1 dyn cm $^{-2}$ to calculate
the mass loss from the surface of the planet. It is unclear what
the suitable lower boundary is? Here we have also computed a model
in which the lower boundary is set to P$_{R_{p}}$=10 dyn cm $^{-2}$.
As expected, the mass loss rates are higher than those in the cases of
P$_{R_{p}}$=1 dyn cm $^{-2}$. Specially, the mass loss rates
enhance 40-70\% for the all cases (see Table 1.).
This hints that the mass loss rates are very sensitive to lower boundary
conditions, and same conclusion has also been found by Penz et al. (2008).

In addition, a possible cause that can lead to the reduction of mass loss rate is the
day-night temperature contrasts at the lower boundary. Theoretical
studies have shown that the day-night temperature contrasts (300K or
more) are likely to occur in the planetary photospheres (Showman \&
Guillot 2002). These predictions are also verified by the observed
surface temperature distribution of HD 189733b (Knutson et al.
2007).

In the above calculations, we used the same temperature over the lower
boundary. In order to further discuss the influence of day-night
temperature contrasts, we set the temperature at the lower boundary
to

\begin{equation} \label{eq:2}
T(R_{p},\theta)= \left\{
\begin{array}{ll}
1500K, (\theta \leqslant 90^{\circ}) \\
1500K \max(0.5,\sin \theta), (\theta > 90^{\circ}).
\end{array}
\right.
\end{equation}

Our results did not show more significant anisotropy for the winds
(These results are not presented in this paper.), and the predicted mass loss rates,
4.05$\times$10$^9$ g s$^{-1}$, is slightly lower than that of model 9 (see Table 1).

\subsection{Dependence of $\dot{M}$ on Stellar Activities}

The Swift XRT light curve of HD189733 shows that the escaping
hydrogen is observed when the star exhibits a bright flare that occurrs about 8
hours before the planetary transit (Lecavelier des Etangs et al. 2012), which
implies a possible correlation between the high level of stellar activity and the atmospheric escape of
hydrogen. Thus, we also calculated the cases with high level of stellar activity (with a solar proxy $P_{10.7}=200$).
For the high activity, the whole EUV flux is about 2.5 times of the low activity.
Our calculations show that the mass loss rates produced by high activity are a factor of 5-10 higher than those
of low activity, and the temperatures predicted are also higher.
By comparing the structures of wind, we found that the model with higher stellar activity
predicted higher radial velocity, and the ratios between meridional and radial velocity are
lower. Evidently, higher activities supply more UV radiations,
especially in the band of higher energy (note that the increases of photon energy are
not linear with frequency between the low and high stellar activity). More high energy photons mean that the deeper base of the
atmosphere can be heated, and the hydrodynamic escape parameter, $\lambda=\frac{GM_{p}\mu}{R_{p}kT}$ ($\mu$ is mean mass per particle),
can decrease to a lower value at deeper atmosphere.
Thus gas pressure drives a faster expanding atmosphere
and partly counteracts the effect of the optical depth. It explains the reason why the ratio of mass loss rate
is only 2.4 for model 8 and 12.

\subsection{Comparison with Observations}
An interesting issue is which mass loss rate predicted by our model is supported
by the observations. Vidal-Madjar et al. (2003) attributed the excess absorption of HD 209458 in Ly $\alpha$ to
a mass loss of 10$^{10}$ g s$^{-1}$. Subsequent 1D models also support this.
However, the influences of two-dimension and charge
exchange between the stellar wind and the planetary escaping
exosphere were neglected by all the 1D models.
Recently, Lecavelier des Etangs et al. (2012) updated the
mass loss rate to 10$^{9}$ g s$^{-1}$ for the atmospheric escape of
HD 189733b though the previous mass loss rate requested in their model is in the order
of magnitude of 10$^{9}$- 10$^{11}$ g s$^{-1}$.
Moreover, Holmstr\"{o}m
et al. (2008) found that a mass loss rate of 7 $\times$ 10$^{8}$ g s$^{-1}$
can explain the observations of HD 209458b when the charge
exchange is included.
We must be careful in the interpretation of mass loss rates because only atomic
hydrogen contributes to the excess absorption of Ly $\alpha$.
In our models, the  mass loss rates of atomic hydrogen are 3.78 $\times$ 10$^{8}$ g s$^{-1}$,
9.85 $\times$ 10$^{8}$ g s$^{-1}$, 7.60 $\times$ 10$^{9}$ g s$^{-1}$ and
1.45 $\times$ 10$^{10}$ g s$^{-1}$ for models 9-12, respectively.
If the charge exchange indeed results in the excess absorption of Ly $\alpha$,
our results favor the low activity for HD 209458 when it is transited by HD 209458b. However,
the possibility that planetary origin hydrogen result in excess absorption of Ly $\alpha$
cannot be excluded because Koskinen et al. (2010) explained this by solely absorption
in the upper atmosphere based on the 1D assumption. Their model denotes the averaged property of the atmosphere,
but is very different with our 2D hydrodynamic model with
consistent radiative transfer.
The differences appear in several aspects. 
First, they estimated a mass loss rate of 1-10$\times$10$^{10}$ g s$^{-1}$,
which is higher than our result. Second, in order to fit the observations, 
they estimated a number density of 2.6$\times$ 10$^{7}$ cm$^{-3}$ for 
hydrogen atoms at r=2.9R$_{p}$. 
However, we found that the number densities of hydrogen atoms in our model are about 10$^5$-10$^6$ cm$^{-3}$ at r=2.9R$_{p}$ (depend on $\theta$),
and the number density distributions of hydrogen atoms show clear variations from the day-side to night-side (see Figure. 13). 
Third, they assumed an ionization of 100\% at r=2.9R$_{p}$.
Our results show that the ionization structures of hydrogen strongly depend on $\theta$ for the case of tidal-locking.
For example, $\frac{n_{h}}{n_{p}}$=0.16, 0.19 and 1.08 at $\theta$=0, $\pi$/2 and $\pi$ at r=2.9R$_{p}$.
However, our models predicted the similar temperature distributions (T$\approx$7000-11000K)
below r=2.9R$_{p}$ as estimated by their models.
Thus, one-dimensional models are not accurate enough in calculating the HI transit depth.
Our results can be applied further to check if the upper atmosphere contributes
significant absorption.

\begin{table*}
 \centering
 \begin{minipage}{140mm}
  \caption{The mass loss rates for HD 209458b.}
  \begin{tabular}{cccccccc}
  \hline
   Model & P$_{10.7}\footnote{The Solar activity proxy.}$ & P(R$_{p})\footnote{The pressure at the lower boundary.}$ & Hydrodynamics & Radiative transfer & $\dot{M}$\\
         &            & (dyn cm$^{-2}$)&            &                    &($\times$ 10$^{10}$g s$^{-1}$)\\
 \hline
 1 & 80 & 1 & 1D & 1D & 3.22 \\
 2 & 80 & 10 & 1D & 1D & 4.78   \\
 3 & 200 & 1 & 1D & 1D & 15.4   \\
 4 & 200 & 10 & 1D & 1D  & 20.2  \\
 \hline
 5 & 80 & 1 & 2D & 1D  &1.72  \\
 6 & 80 & 10 & 2D & 1D  &2.60  \\
 7 & 200 & 1 & 2D & 1D  &12.1  \\
 8 & 200 & 10 & 2D & 1D  &15.6 \\
 \hline
 9 & 80 & 1 & 2D & 2D & 0.43  \\
 10 & 80 & 10 & 2D & 2D & 0.77   \\
 11 & 200 & 1 & 2D & 2D  & 4.50 \\
 12 & 200 & 10 & 2D & 2D  &6.55 \\
\hline
\end{tabular}

\end{minipage}
\end{table*}

\begin{figure*}
\epsscale{1.00} \plotone{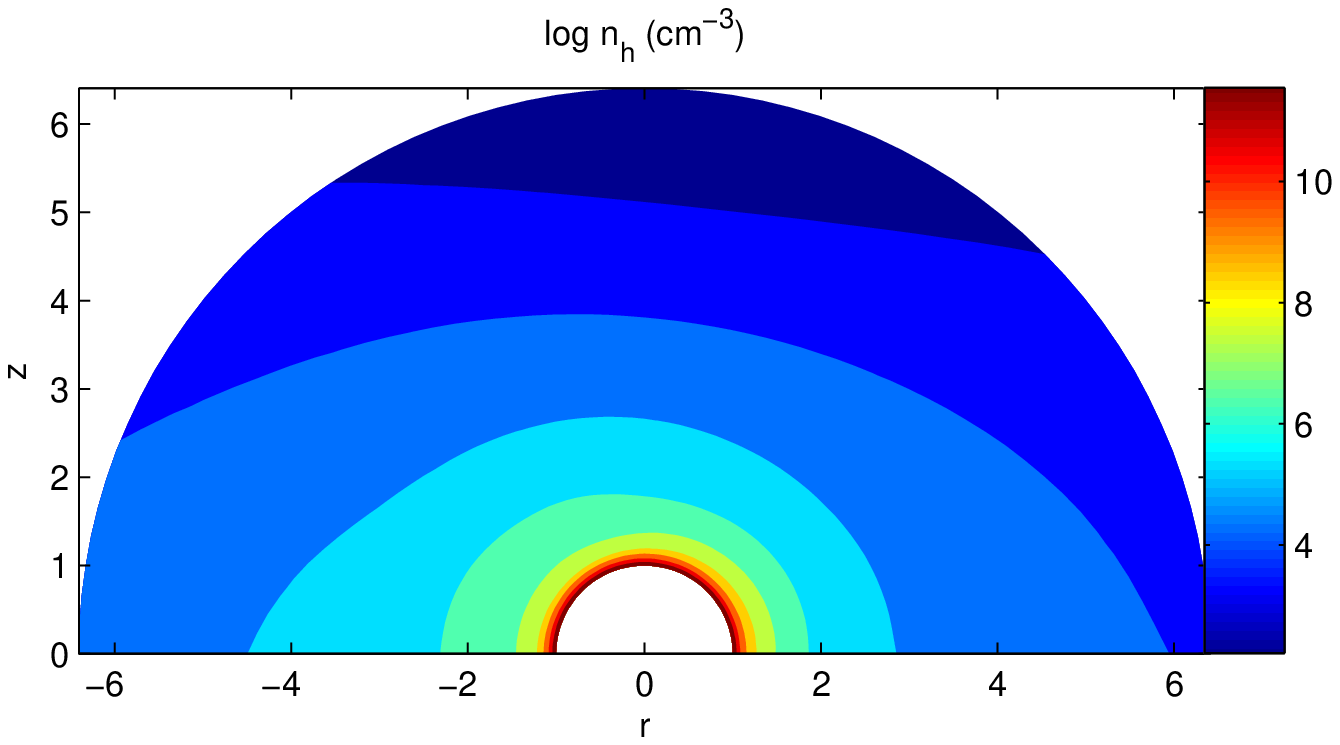}  \caption{The number density distributions of hydrogen atoms in the case of
the two-dimensional hydrodynamic model
with two-dimensional radiative transfer.}
\end{figure*}

\section{CONCLUSIONS}
HST observations of HD 209458b and HD 189733b in the UV band have
revealed the signatures of atmosphere escape. Theoretical progress is
following the observations, for example, reasonable mass
loss rates have been predicted by several one-dimensional hydrodynamic
models (Yelle 2004; Tian et al. 2005; Garc\'{\i}a Mu\~{n}oz 2007;
Penz 2008; Murray-Clay et al. 2009; Guo 2011). Two-dimensional hydrodynamic model have also
revealed a possible anisotropy in the planetary wind (Stone \& Proge 2009).
However, there exists still a few key problems to be solved, such as
the proper descriptions to radiative transfer and charge-exchange reactions
between the planetary and stellar wind
particles. They play an important role
in fully explaining the observations.

In this paper, we presented a two-dimensional multi-fluid hydrodynamic model by
incorporating the two-dimensional radiative transfer.
From this work, our conclusions are the following.

1. In the assumption of non-tidal locking,
   the planetary winds are not spherically symmetric. Tidal
   forces can result in significant horizontal
   movements in the planetary winds and lower density and radial velocity
   near the polar regions so that the mass is almost lost fully from
   the low-latitude zones.

2. In the case of tidal locking, we found that the
   wind is controlled by meridional velocity in the regions of r/R$_{p}\sim 1.5$.
   The large meridional
   velocities transfer the mass and energy from the day side to the night side at the base of the wind.
   But the extent decreases with an increase in the radius due to the decrease of the optical depth.
   At large radii, the wind is siginificant difference with that of one-dimensional radiative transfer.
   Due to the declines of density, the mass loss rate is also lower than
   that of non-tidal locking.

3. The mass loss rates are affected by both two-dimensional radiative transfer
   and hydrodynamics. Comparing with the results of one-dimensional hydrodynamic models,
   the models of two-dimensional hydrodynamics predicted the mass loss rates of
   1.72 and 0.43 $\times$ 10$^{10}$ g s$^{-1}$
   for one-dimensional and
   two-dimensional radiative transfer, which are a factor of 2 and 7 smaller than that
   of 1D hydrodynamic model.

\section*{Acknowledgments}

The author are grateful to the referee for comments that led to substantive improvements in this work.
This work was supported by National Natural Science Foundation of
China (Nos.10803018 and 11273054).

\clearpage

\end{document}